\documentclass{aastex} 
\begin{document}
\received{}
\accepted{}
\revised{}
\slugcomment{Draft, 07/16/01 -- v12.2}
\shortauthors{Pritzl et al.}
\shorttitle{NGC~6441}
\title{Variable Stars in the Unusual, Metal-Rich Globular Cluster 
 NGC~6441} 
\author{Barton J. Pritzl\altaffilmark{1,2} and Horace A. Smith}  
\affil{Dept. of Physics and Astronomy, Michigan State University, 
East Lansing, MI 48824\\e-mail: pritzl@pa.msu.edu, smith@pa.msu.edu}
\author{M\'arcio Catelan\altaffilmark{3}} 
\affil{University of Virginia, Department of Astronomy, P.O.~Box~3818, 
Charlottesville, VA 22903-0818\\e-mail: catelan@virginia.edu}
\and
\author{Allen V. Sweigart}
\affil{NASA Goddard Space Flight Center, Laboratory for Astronomy and 
Solar Physics, Code~681, Greenbelt, MD 20771\\e-mail: 
sweigart@bach.gsfc.nasa.gov}

\altaffiltext{1}{Visiting Astronomer, Cerro Tololo Inter-American
Observatory, National Optical Astronomy Observatories, which is
operated by AURA, Inc., under cooperative agreement with the
National Science Foundation.}
\altaffiltext{2}{Current address: National Optical Astronomy Observatories, 
P.O. Box 26732, Tucson, AZ 85726, email: pritzl@noao.edu}
\altaffiltext{3}{Hubble Fellow.}

\newpage

\begin{abstract}

We have undertaken a search for variable stars in the metal-rich
globular cluster NGC~6441 using time-series $BV$ photometry.  The 
total number of variables found near NGC~6441 has been increased to $\sim 104$, 
with 48 new variables being found in this survey.  A significant number
of the variables are RR Lyrae stars ($\sim 46$), most of which are probable
cluster members.  As was noted by Layden et al. (1999), the 
periods of the fundamental mode RR Lyrae are unusually long compared to field 
stars of similar metallicity.  The existence of these long period RRab
stars is consistent with Sweigart \& Catelan's (1998) prediction that
the horizontal branch of NGC~6441 is unusually bright.  This result implies 
that the metallicity-luminosity relationship for RR Lyrae stars is not 
universal.  We discuss the difficulty 
in determining the Oosterhoff classification of NGC~6441 due to the unusual 
nature of its RR Lyrae.  A number of ab-type RR Lyrae are found to be both 
brighter and redder than the other probable RRab found along the horizontal 
branch, which may be a result of blending with stars of redder color.  
A smaller than usual gap is found between 
the shortest period fundamental mode and the longest period 
first-overtone mode RR Lyrae.   We determine the reddening of the cluster to 
be $E(\bv) = 0.51 \pm 0.02$~mag, with substantial differential reddening
across the face of the cluster.  The mean $V$ magnitude of the RR Lyrae is 
found to be $17.51 \pm 0.02$ mag resulting in a distance of 10.4 to 
11.9~kpc, for a range of assumed values of $\langle M_V \rangle$ for RR Lyrae 
stars.  The possibility that stars 
in NGC~6441 may span a range in [Fe/H] is also discussed.

\end{abstract} 


\newpage

\section{Introduction}

Armandroff \& Zinn (1988) determined the metal abundance of
NGC~6441 to be [Fe/H] = -0.53 $\pm$ 0.11, making it slightly
more metal-rich than 47 Tucanae on their metallicity scale.
One might therefore have expected that the color-magnitude
diagram (CMD) of NGC~6441 would show the stubby red horizontal branch (HB)
typical of metal-rich globular clusters such as 47 Tuc.
The CMD of NGC~6441 does indeed show a
strong, red HB component, but Rich et al. (1997), using {\it Hubble
Space Telescope} (HST) observations, discovered that the CMD of
NGC~6441 also has a pronounced blue HB component, which stretches
across the location of the instability strip. Not only does the HB have 
a blue component, 
but it also slopes upward as one goes blueward in a ($V$, $\bv$) CMD.  
NGC~6441 is not, however, unique 
in these characteristics:  Rich et al. showed that the CMD of the 
relatively metal-rich globular cluster 
NGC~6388 has a similar HB morphology.  It has long been known that the 
HB morphology 
of globular clusters does not correlate perfectly with [Fe/H] and that
at least one parameter besides [Fe/H] is needed to
account for this (Sandage \& Wildey 1967; van den Bergh 1967). 
NGC~6441 and NGC~6388 are the only metal-rich globular clusters
known to exhibit this second parameter effect. 

Sweigart \& Catelan (1998) constructed CMD simulations of metal-rich 
globular clusters to test various theoretical scenarios which might 
account for the unusual HB morphologies in NGC~6441 and NGC~6388.  They 
found that the upward slope of the HBs could not be explained by
differences in either cluster age, mass loss on the red giant branch (RGB),
or differential reddening.  Layden et al. (1999) would later,
on observational grounds, also rule out the hypothesis of
differential reddening.  Sweigart \& Catelan did arrive at
three theoretical scenarios for explaining the HB
morphology: (1) a high cluster helium abundance; (2) a rotation
scenario in which rotation during the RGB
phase increases the HB core mass; and (3) a mixing scenario,
in which deep mixing on the RGB enhances the
envelope helium abundance.  All three of these scenarios
predict that any RR Lyrae stars (RRLs) in NGC~6388 and NGC~6441 should
be anomalously bright.  Although at the time little was known
about RR Lyrae stars in NGC~6441, the few RR Lyrae stars
known in NGC~6388 had properties which were consistent with an 
unusually high luminosity (ref. Fig. 3, Sweigart \& Catelan 1998). 

Sweigart (1999) suggested an additional scenario: Some stars might form 
dust grains which are then removed from the envelope by radiation 
pressure at the tip of the RGB, leaving behind metal-depleted gas.  The overall
metal abundance of the envelope might then be reduced if the metal-depleted
gas can be convectively mixed throughout the envelope.  One would then
expect an HB star that has undergone such metal-depletion to be
both bluer and brighter than would otherwise be the case.  As a variation 
of this metal-depletion scenario, Sweigart (2001) suggested that the 
unusual HB morphology in NGC~6441 and NGC~6388 might also be due to 
an intrinsic spread in metallicity among the stars in these clusters, a 
possibility first noted by Piotto et al. (1997).  Detailed stellar 
evolution calculations have shown that a metallicity spread can produce 
upward sloping HBs similar to those observed in these clusters 
(Sweigart 2001).  Under this scenario NGC~6441, along with NGC~6388, 
would be metal-rich analogs of $\omega$~Centauri, the only Galactic 
globular cluster known to contain a spread in metallicity.

Layden et al. (1999) used ground-based $V$ and $I$ photometry
of NGC~6441 to study its CMD and its
variable star population.  They discovered about 50 new
variable stars in the vicinity of NGC~6441, including 11 RRLs which they 
believed to be probable members of the cluster and 9 other suspected cluster 
variables, 6 of which were candidate RRLs.
The RRab stars which they believed to be probable members had unusually 
long periods.  Their locations in the period-amplitude diagram
(ref. Figure 9, Layden et al. 1999) were consistent with the models
of Sweigart \& Catelan (1998), implying that the RRLs were
indeed unusually bright. Layden et al. also found a value of 1.6
for the ratio $R$ of HB stars to RGB stars, implying that
the helium abundance of NGC~6441 is not extraordinarily high.  Thus, 
the high helium scenario seems to be effectively ruled out. 

This paper reports new $B$ and $V$ photometry of NGC~6441 which has
led to the discovery of additional variable stars.  Preliminary results 
from these observations have already been used to argue that the 
RRLs in NGC~6441 are unusually bright for the cluster metallicity (Pritzl et 
al. 2000).  Here we present the results of the new study in detail and 
call attention to several unusual properties of RRLs in NGC~6441.  The 
results for NGC~6388 will be reported in a separate paper (Pritzl 
et al. 2001).

\section{Observations and Reductions}

Time series observations of NGC~6441 were obtained at the 
0.9 m telescope at Cerro Tololo Inter-American Observatory using the 
Tek 2K No. 3 CCD detector with a field size of 13.5 arcmin per side.
Time series observations of NGC 6441 were obtained on the UT dates 
of 1998/May/26-29 and 1998/June/1-4. Exposures of 
600 seconds were obtained in both $B$ and $V$ filters.  The seeing 
ranged from 1.1 to 2.5 arcsec, with a typical seeing of 1.4 arcsec.  
Images were centered 4 arcmin east of
the cluster center, to avoid including a bright foreground star
within the field.    

Twilight sky flats were taken on nights 1, 2, 4, 5, and 6.  The sky flats 
were examined for any nightly variations and no significant variations were 
detected.  The sky flats were then combined into one master sky flat that 
was used for each frame.  Corrections for bias and flat fielding were
done using conventional {\sc iraf} routines.  

A total of 23 photometric standards from Landolt (1992) and 
Graham (1982) were observed on June 1, 2, and 3.  They were typically observed 
four times during the night.  These 
primary standard stars spanned a color range from $\bv = 0.024$ to 2.326~mag, 
adequate to cover the color range of the reddened stars in NGC~6441.  Primary 
standards were observed between airmasses 1.035 and 1.392.  According to the 
CTIO sky condition archive 
(http://www.ctio.noao.edu/site/phot/sky\_conditions.html), only the night of 
June 1 was photometric, but standards observed on the nonphotometric nights 
were incorporated in the reductions using the ``cloudy" night reduction 
routines created by Peter Stetson (private communications).  This allowed us 
to make use of a broader range of colors from the primary standards observed 
on the other nights to better define the color terms in the transformation 
equation.

Photometry of stars in the NGC~6441 field was obtained using the
standalone reduction packages created by Stetson (1994): {\sc daophot ii},
{\sc allstar}, and {\sc allframe}.  Point-spread functions (PSF) were 
calculated for each frame from a set of 80 bright, uncrowded stars, using the 
variable PSF option.  Instrumental magnitudes $v$ and $b$ for the NGC~6441 
stars were transformed to Johnson $V$ and $B$ magnitudes using Stetson's 
{\sc trial} package. Thirty nine local
standards within the NGC~6441 field were used to set the frame-by-frame 
zero-points for the cluster observations.  Because NGC 6441 was 
observed to higher airmass than the Landolt (1992) and Graham (1982) 
standards, the local 
standards were also used to check the adopted values of the extinction 
coefficients for the night of June 1.  The observations of the local 
standards confirmed the values of the extinction coefficients determined 
from the primary standards.  Transformation equations derived from the 
standard stars had the form:  

\begin{equation} 
v = V - 0.006 \, (\bv) + 0.159 \, (X-1.25) + C_V 
\end{equation} 

\begin{equation} 
b = B + 0.105 \, (\bv) + 0.243 \, (X-1.25) + C_B  
\end{equation} 

\noindent where $X$ is the airmass and $C_V$ and $C_B$ are the 
zero point shifts for their respective filters.
Comparing the transformed magnitudes of the standard stars with the
values given by Graham and Landolt, we find rms residuals of 0.018 magnitudes in
$V$ and 0.006 magnitudes in $B$.

The photometry of stars in the NGC~6441 field could be compared with 
photometry in three earlier studies, as shown in Table~1.  First, our
$B$ and $V$ photometry was compared with the HST $B$, $V$ photometry obtained
by Rich et al. (1997).  Unfortunately, 
even the outermost stars in the field of view of WFPC2 tended to be
crowded on the images obtained at CTIO.  
The comparison with Rich et al. is therefore 
based upon only 13 stars for which the crowding on our images,
although still significant, is small enough to allow a meaningful
comparison.  Our ground-based photometry is brighter in $V$ by about
0.04 mag.  No significant difference in $B$ is evident.  

A larger number of comparisons could be made with the photoelectric 
and photographic photometry of Hesser \& Hartwick (1976).  In Figure 1 
we show the differences between our results and the local standards 
established by Hesser \& Hartwick.  At low $\bv$, there is some scatter, 
but the mean differences are small.  The results are more discordant for 
the reddest stars.  Excluding stars with $\bv > 2.0$~mag, we obtained 
the results listed in Table~1.  Our mean $V$ and $B$ values are slightly 
brighter than those of Hesser \& Hartwick by an amount smaller than 
or comparable to the standard error of the mean.  We are not certain of 
the origin of the greater discrepancy for very red stars, though we note 
that these more discrepant stars tended to be among the fainter as well 
as the redder stars used in the comparison.  Our inclusion of a 
second-order color term in our transformation equations produced no 
significant decrease in the discrepancy at larger $\bv$ values.  It 
is noteworthy that Figure 1 shows no trend with $\bv$ for the bluer 
colors relevant to the region of the RRLs.  No trend with color was 
seen in the comparison of our results with the HST photometry.

Finally, we compared our $V$ magnitudes with the $V$ 
photometry of Layden et al. (1999).  As indicated in Table~1, our $V$ 
photometry and that of Layden et al. are surprisingly discrepant.  We 
have no good explanation for this discrepancy.  The agreement between 
our $V$ photometry and that of Rich et al. (1997) and Hesser 
\& Hartwick (1976) suggests that there is a zero-point error of 
approximately 0.4~mag in the Layden et al. photometry.

\section{Color-Magnitude Diagram} 

The CMD was the critical tool for the initial 
discovery of the unusual nature of NGC~6441.  For this reason we will now 
compare our new CMDs with those presented by Rich et al. (1997) and 
Layden et al. (1999). 

Figure 2 shows four CMDs obtained from our photometry for stars lying at 
different distances from the center of NGC~6441.  A total of 14127 stars 
make up the CMD in Figure 2a.  Only the stars with values of $\chi < 1.5$ 
from Stetson's {\sc trial} program were included.  The strong red component 
of the HB is evident ($V \sim 18$, $\bv \sim 1.4$) as is its blue 
extension.  We also see many features of the field bulge population,
which are particularly prominent since our images 
are shifted off of the cluster center to avoid a foreground star.  The main 
sequence of the field extends up through the cluster's HB from about 
$\bv \sim 1.2$ to $\sim 0.8$~mag and from about $V \sim 20$ to
$\sim 15.5$~mag.  The field red HB clump is found at $V \sim 17$~mag and 
$\bv \sim 1.7$~mag.  The contribution of the bulge stars to the CMD of 
Figure 2a can be seen in Figure 2d, where we have plotted all stars that 
are 6-11 arcmin eastward from the cluster center [the tidal radius of 
NGC~6441 extends to 8.0 arcmin (Harris 1996)].  

One interesting feature on the diagrams of NGC~6441, which can also be 
seen in the CMDs of Rich et al. (1997) and 
Layden et al. (1999), is a small clumping of stars slightly fainter and 
redder than the red HB of NGC~6441 ($V \sim 18.5$, $\bv \sim 1.4$).  
This appears to be the RGB luminosity 
function bump.  Although this bump is difficult to discern in Figure 2, 
its presence in the HST 
CMD makes it likely to be associated with the cluster.  An investigation 
into its properties might constrain the helium content of the 
cluster (Sweigart 1978; Fusi Pecci et al. 1990; Zoccali et al. 1999; 
Bono et al. 2001).

Figure 2b is our closest approximation to the area covered by the CMD 
of Rich 
et al. (1997), including stars within a radius extending out to 
approximately 1.7 arcmin from 
the cluster center.  The morphology of the HB is clearer in this figure 
and Figure 2c.  Not only do we see the sloped nature of the blue 
extension to the HB, but the slope in the red HB is also seen as was first 
noted by Piotto et al. (1997).  The HB slopes from $V\sim 18$~mag, at 
$\bv\sim 1.4$~mag, to $V\sim 17.5$~mag, at $\bv\sim 0.6$~mag. 

Rich et al. (1997) mentioned that there may be a hint for a bimodal 
distribution of the stars on the HB of NGC 6441.  There appears to be 
a gap ranging in color from 1.0 to 0.4 mag in their Figure 1.  Such a 
bimodality is not seen in our CMDs, although field stars and 
differential reddening may be filling in any gap in the HB.  It is 
possible that this gap may be due to the RRL being plotted at their 
instantaneous magnitudes instead of their time averaged magnitude and 
color in the Rich et al. CMD.  Looking 
at our Figure 4, it is seen that the RRL fill in the HB in that region.

\section{Variable Stars} 

\subsection{Discovery of New Variable Stars}

Variable stars were identified in two ways: first, Stetson's {\sc daomaster}
routine was used to compare the rms scatter in our photometric values
to that expected from the photometric errors returned by the {\sc allframe}
program. Second, we applied the variable star search algorithm
presented by Stetson (1996).  Results from the two approaches were
very similar.

The time coverage of our observations is well suited for the
discovery of short period variability, but not for the detection
of long period variables.  All of the probable short period variable
stars identified by Layden et al. (1999) which were within our field 
were recovered during
our variable star searches.  In addition, 48 probable new variable
stars were detected, along with 6 suspected variables.  In crowded 
regions close to the cluster
center, the $B$ photometry proved superior to the $V$ photometry for
purposes of identifying variable stars, presumably because of the
lesser interference from bright red giant stars.  Finding
information for the new variable stars is given in Table~2, where 
$X$,~$Y$ 
are the coordinates of the variables on the CCD [the cluster center is 
assumed to be at (1635,1051)] and $\Delta\alpha$,~$\Delta\delta$ are the 
differences in right ascension and declination from the cluster center 
(in arcsec).  Finding charts for the variables, excluding the long period 
variables, can be seen in Figure 3.

\subsection{RR Lyrae stars} 

The number of probable RRL stars in the NGC~6441 field has been 
increased from 11 to 46.  The location of these stars within the CMD 
is shown in Figure 4.  All previously known cluster RRL have been 
rediscovered except for Layden et al.'s (1999) 
V36, a field star lying outside 
the field of our observations.  Table~3 lists the mean properties of the 
individual RRL stars found in this survey, along with one $\delta$ Scuti or 
SX Phoenicis star.  All periods were found 
using the phase dispersion minimization program in {\sc iraf}.  The 
accuracy of these periods is $\pm 0.001$~d to $\pm 0.002$~d, depending on 
the scatter and completeness of the light curve.  The periods 
determined for the known RRL are in good agreement with those found by 
Layden et al.  Magnitude weighted and luminosity weighted mean magnitudes 
were calculated using spline fits to the observations.  The listed 
$\langle V \rangle$ values are luminosity weighted, but the colors are 
magnitude weighted, $(\bv)_{\rm mag}$.  Light curves and photometry 
for the variable stars in $V$ and $B$ are given in Figure 5, Table~4 and 
Table~5, respectively.  
The mean level of the NGC~6441 HB determined from the probable 
RRL members, excluding the brighter and redder RRab (see \S4.4) and 
RRL with uncertain classification, is $V = 17.51 \pm 0.02$~mag. 

As was noted by Layden et al. (1999), it can occasionally be difficult 
to distinguish RRc variables from eclipsing binary stars which have 
periods twice as long.  This is particularly true for a cluster such as 
NGC~6441, in which a significant and variable reddening makes the 
precise location of a variable star in the CMD an 
uncertain guide as to the type of the variable.  Although not 
always decisive, inspection of the Fourier decompositon parameters of 
the light curves can be an aid in classifying variables, and in 
distinguishing RRab from RRc variables.

Fourier decompositions of the light curves were done fitting to an equation 
of the form: 

\begin{equation} 
mag = A_0 + \sum A_j \, \cos(j\omega t + \phi_j). 
\end{equation} 

\noindent 
Plotting the Fourier parameters $A_{21}$ vs. $\phi_{21}$ gives a clear 
distinction between RRL types (e.g. Clement \& Shelton 1997), where 
$A_{21}=A_2/A_1$ and $\phi_{21}=\phi_2-2\phi_1$.  Figure 6 
shows a $A_{21}$ vs. $\phi_{21}$ plot for the probable RRL variables 
in the NGC~6441 field which have clean light curves.  The data are 
listed in Table~6.  The probable RRL fall into areas of the diagram where 
RRab and RRc stars would indeed be expected.  A clear break 
between the RRL types can be seen at $A_{21}$ of 0.3 as was originally 
shown by Simon \& Teays (1982), with the RRab falling at values greater 
than this and the RRc below.  It should be noted that V49 falls in the 
same region as the other RRc variables furthering the case for its 
reclassification (see \S4.3).  We also see how the field star V54 stands 
apart from the cluster variables. 

Using these RRL with quality light curves as a reference, we are able 
to analyze the other variables found in the field of NGC~6441.  Table~7 lists 
the Fourier values for the contact binaries and the RRL with poorer light 
curves and uncertain 
membership or classification.  It is clear that the contact binaries have 
values of $A_{21} > 1.0$.  This provides a good distinction between the RRL 
and the contact binaries.  There is still the question of how to determine if 
a variable is a RRc star or a contact binary star with a period twice as 
long.  To test this we wanted to see how the Fourier parameters of a known 
RRc star would compare to those of known contact binaries if it was analyzed 
at the longer period that it would have were it a contact binary.  We 
performed a Fourier analysis on V70, which is clearly 
a RRc variable and fits in well with the other RRc in the $A_{21}$ vs. 
$\phi_{21}$ plot, at the longer period of 0.852~d.  The resulting $A_{21}$ 
value, 25.90, would indicate that 
it was a binary star.  In order to test how contact binary stars compare to 
known RRc stars, we also analyzed two stars that 
are clearly contact binary stars, V86 and V88, at the shorter periods that 
they would have 
if they were RRc stars.  The data is shown in Table~8.  Comparing these results 
to the RRc in Figure~6, the $A_{21}$ values are a little high and the 
$\phi_{21}$ values are low.  Although more tests should be done to verify this, 
it appears that if one 
analyzes a star at a RRc-type period, a true RRc star will fall among the 
other RRc stars in the $A_{21}$ vs. $\phi_{21}$ plot, while a binary star will 
be offset to low $\phi_{21}$ values and slightly higher $A_{21}$ values.
For comparison, as stated before, the Fourier values for V49 imply 
that it is a RRc variable rather than a contact binary.  On the other hand, 
the values for V81 at the shorter RRc-like period imply that it may indeed be 
better classified as a contact binary.

The period-amplitude diagrams for this cluster in $B$ and $V$ are seen in Figure 
7.  There are a few RRab variables whose amplitudes are low for 
their period.  Some of these are probably attributable to noisy photometry 
(V46, V52, V53, and V55),  
while others 
are due to possible blending effects (see \S4.4).  
The Blazhko effect can also reduce the amplitudes of RRab stars, but our 
observations do not extend over a long enough time interval to test for the 
presence of this effect.  
Another striking 
feature is the lack of a significant gap between the shortest period RRab 
and the longest period RRc (see \S5.2 and \S5.4).  

\subsection{Notes on Individual RR Lyrae} 

\indent 
V45 -  V45 appears to stand out from the other cluster variables.  The 
shape of the light curve indicates that the variable is of ab type.  
Although our data have a gap in the light curve near
minimum making its magnitude and color uncertain, we were able to determine 
a period similar to that in Layden et al. (1999), whose
phase coverage is more complete.  Layden et al. list the $V$
amplitude of the star as 0.73~mag.  We estimate the amplitude to be about 
0.85~mag from
our data.  When we place this variable in the period-amplitude diagram for 
the cluster
(Figure 7), it clearly stands apart from the variables we feel certain are
cluster members.  It is our belief that V45 is a field RRL that happens
to fall at nearly the same distance as the cluster.

\indent
V49 - V49 was classified by Layden and collaborators as a possible 
detached binary with a period of 1.010~d.
Our data indicate that this variable may instead be classified as an RRc-type
variable with a period one third as long.  At a period of 0.335~d, our data 
for one night ($\sim 8$~h) nearly complete one cycle.  When the  
Layden et al. (1999) observations are fit to a 0.335~d period, we get 
a light curve similar to our own, although with some scatter.   
V49 lies toward the outer parts of NGC~6441.  The color of the 
star fits well with the other cluster RRc, but
it is slightly brighter.  As stated in \S4.2, the Fourier values 
for this star place it among the other RRc stars.  We are
unable to determine what, if any, reddening effects may be responsible for the
difference in brightness.  Although we believe that V49 is more
likely to be an RRc than a detached binary, more photometry of this
star would be useful in making a definite determination.   

\indent 
V52 - This variable has a period, magnitude, and color that would classify it 
as being an ab-type RRL.  Yet, it is interesting to note that its light curve 
has a more sinusoidal shape.  To illustrate the difference, it is useful to 
compare the light curve of V52 with those of V46, whose period is slightly 
longer, and V53, whose period is similar to V52.  Both V46 and V53 exhibit 
a sharper rise in light to maximum, whereas V52 has a more gentle slope.

\indent 
V54 - This variable is an RRab with a clean light curve.  It is $\sim 0.9$~mag 
brighter than other cluster RRab variables.  In addition, the location of 
V54 far from the cluster center indicates that it is a likely field 
variable.  V54 may also be intrinsically fainter than the cluster RRab 
given its position in the period-amplitude diagram (Figure 7).

\indent 
V64 - This star is blended with a close neighbor, only 
1.3 arcsec away.  The RR Lyrae-like periodicity shows up in the
photometry of both stars, indicating that the photometry of both is
probably affected by the blending. Therefore, it is likely that only one of 
the two stars is actually a variable.  The $B$ light curves for 
each star have scatter, with the amplitude of V64 being 0.95 mag and that of
the neighbor being 0.60 mag.  The $V$ curves show a lot of scatter with amplitudes 
around 0.30 mag.  The mean magnitudes for the companion star are $\langle{V}
\rangle=16.971$~mag and $(\bv)_{\rm mag}=1.378$~mag.  

\indent 
V67 - This variable is found only $\sim 3.5$~arcsec east from a much
brighter star.  Therefore blending was a problem for some of the nights 
with poorer seeing.  These data points were removed from the light curve.  
Although its color is similar to the other RRab 
stars believed to be on the NGC~6441 HB, it is slightly 
brighter.  This may be a consequence of its proximity to the bright star.  
It should be noted, however, that the ratio of $B$ to $V$ amplitude is not 
unusual.  Its large distance away from the cluster center could 
indicate that V67 is a field star.  Unlike V45 and V54, V67 falls along with 
the other probable cluster members in the period-amplitude diagram (Figure 7).

\indent 
V68 - (SV1, Layden et al.) We are still uncertain 
as to how to classify this variable.  It appears as though the scatter 
in its curve comes from blending with a close companion  
star.  Also, coma may be affecting our photometry since V68 is found near 
the edge of the frame.  V68 is unusually bright, and if it is found to be 
an RRc variable, it should be considered to be a member of the field.  

\indent
V69 - (SV2, Layden et al.) Even though it has an 
unusually long period, the light curve shape coupled with the location in the 
CMD indicates that V69 is a cluster RRL of c-type.  (See \S6.2.)  

\indent 
V70 - (SV4, Layden et al.)  The magnitude and color of this variable, along with 
the shape of the light curve, indicate it is a c-type RRL.

\indent 
V71 - (SV5, Layden et al.)  The magnitude and color of this variable, along 
with the shape of the light curve, indicate it is a c-type RRL. 

\indent 
V73 - This variable has an uncertain classification.  In the CMD, it is 
slightly brighter and redder than other RRc variables.  The light curve shape 
is somewhat asymmetric, but there is also scatter 
to indicate that there might be some problem with blending.  The 
distance of V73 from the cluster center is large enough to raise the possibility 
that it may be 
a field star. 

\indent 
V75 - This star falls among the RRc stars in the NGC~6441 CMD, but there is some 
scatter in its curve, especially in $V$, that makes its exact classification 
uncertain.

\indent 
V76 - This is a longer period c-type RRL with a curve similar to that of V69.  
It appears to be fainter and redder compared to other RRc stars on the 
HB, which may be an effect of differential reddening. 

\indent 
V78 - This variable is fainter and redder as compared to the RRc stars on the 
HB, which may be an effect of differential reddening. 

\indent 
V79 - V79 is an RRc star as indicated by its $B$ light curve.  The $V$ light curve 
has a lot of scatter in it, and the $\bv$ color for this variable, which is somewhat 
redder than the other RRc stars, is uncertain for that reason.  

\indent 
V80 - (SV3, Layden et al.) Our light curves for this star show that it is 
better classified as a binary rather than an RRc star.  The location of this 
star in the CMD, which puts it in the vicinity of the RRab stars, also indicates 
that the star is not a c-type RRL.

\indent 
V81 - A probable eclipsing binary star, but the phase coverage is not 
complete when plotting the longer period of 0.856 d making the 
classification uncertain. 

\indent 
V84 - This star appears to be of c-type.  It is very blue as compared 
to the cluster RRc.  There is a lot of scatter in the 
curves, especially in the $V$ light curve.  The mean $\bv$ color is
probably unreliable.
 
\indent 
V93 - The scatter found in the light curve of this variable makes it difficult 
to classify.  It has a slightly asymmetric $V$ light curve.  The placement 
of V93 along with the other RRc variables along the horizontal branch suggests 
that it is a RRc variable.  

\indent 
V94 - This variable falls along the horizontal branch, although it is slightly 
redder than most of the RRc variables.  The light curve shows an unusual shape 
having a longer than usual rise time.  The precise classification of this 
variable is uncertain.  We were unable to test if this star is a 
RRc star or a contact binary as in \S4.2 due to the scatter in its 
light curve.
 
\indent 
V95 - From the shape of the curve, the location in the CMD, and its period,  
this star is a foreground $\delta$~Scuti or SX Phoenicis star. 

\indent 
V96 - The somewhat asymmetric light curve and period of this variable indicate 
that it could be an RRab variable.  It is difficult to make an exact 
determination  
of the variable type due to scatter found in the light curve and a gap present 
during the rise in the light curve.  The location of this variable, slightly 
fainter and redder than other RRab variables, may be an effect of 
differential reddening.  

\indent 
V97 - The period found for this variable is similar to those found for other 
RRab variables of NGC~6441.  The $B$ data show definite variability 
with a period of 0.844~d, 
while the $V$ data show large scatter.  We have not given it a definitive 
classification since we see no clear minimum in the light curve.  

\indent 
V102 - The classification of this star is uncertain.  Although the $B$ 
light curve looks to be that of a c-type RRL, V102 is much 
brighter and redder than the other c-type RRL found in NGC~6441.  
The $V$ light curve has more scatter in it than the $B$ light curve, 
implying, as with the red RRab, that blending may be the cause of the 
difference (see \S4.4).  The $V$ amplitude for this star, as 
is, would be at most $\sim 0.1$~mag.   

\indent 
SV6, SV7, \& SV9 - Two of these suspected variables, SV6 and SV7, from Layden 
et al. (1999) did not show any variation in our data.  Layden and collaborators 
designate these stars as possible long period variables (LPVs).  
Since our survey was not geared to  
search for LPVs, these stars may indeed be varying over a larger time scale 
than we were able to sample.  
SV9 was not in our field of view.  

\subsection{``Red" RR Lyrae}

Layden et al. (1999) noted that V41 and V44 stood apart from 
the other cluster RRL in that they were both brighter and 
redder than the expected red boundary of the RRL instability strip.  
With the increased number of RRL found in this survey, we also 
found an increased number of these unusual RRL.  V62 and V65 are 
both brighter by approximately 0.61~mag and redder by approximately 0.25~mag  
in $\bv$ than typical NGC~6441 RRab stars.  

Layden et al. (1999) suggested that these stars were variables with 
unresolved red stars contaminating their photometry.  This seems the 
most likely explanation.  NGC~6441 has a 
very high central stellar density, indicating that crowding effects are 
highly likely.  A consequence of blending with a red star would be an 
unusually high ratio of the $B$ to $V$ amplitude.  This indeed seems to be 
the case for V41, V44, V62, and V65.  Several other possible ``red"
RR Lyrae stars were noted, but they all had large scatter in their
$V$ light curves.   
It should also be noted that the $V$ light curves of these 
variables tended to exhibit a higher scatter 
than the $B$ light curves.  If the unusual color of these
stars is to be explained by unresolved companions, it is perhaps 
unexpected that all four such stars are of type ab and none of type c, 
with the possible exception of V102.
On the other hand, it would be easier to discover 
blended variables with larger amplitude, all else being equal, which
might favor blends involving RRab stars. 

We attempted to search the Rich et al. (1997) images for the 
``red" RRL and other RRL whose $V$ light curves had large scatter
to investigate if there were any nearby neighbors.  We only found 8 variables 
that were within the field of the HST images.  V56, V63, V64, V65, V75, 
V84, and V102 all had from 1 to 3 stars within a 0.5 arcsec by 0.5 arcsec area 
centered on each variable.  V66 had a neighboring star within a radius of 1.2 
arcsec.  Only V97 did not appear to have a neighboring star.  This indicates 
that Layden et al.'s (1999) assertion that these variables 
are having their photometry contaminated by unresolvable stars in the 
ground-based images may be correct.

\subsection{Reddening} 

Sturch (1966) found that near minimum light the blanketing-corrected and 
reddening-corrected color were a function only of period, $P$, for RRab stars. 
The observed color during this phase could therefore be used in
determining the reddening of an RRab star. 
Blanco (1992) 
modified Sturch's procedure by incorporating the metallicity indicator 
$\Delta\,S$.  He found 

\begin{equation} 
E(\bv) = \langle{\bv}\rangle_{\phi(0.5-0.8)}+0.0122\,\Delta\,S-
0.00045\,(\Delta\,S)^2-0.185\,P-0.356. 
\end{equation} 

To infer $\Delta\,S$ for the NGC~6441 variables, we used two different methods.  
The calibration of 
Blanco (1992), which makes use of abundances from high resolution spectra of 
RR Lyrae stars, gives: 

\begin{equation} 
{\rm [Fe/H]} = -0.02(\pm{0.34})-0.18(\pm{0.05})\,\Delta\,S. 
\end{equation}

\noindent
Suntzeff et al. (1991) based their alternative calibration upon the globular 
cluster metallicity scale adopted by Zinn \& West (1984).  They found: 

\begin{equation}
{\rm [Fe/H]} = -0.408-0.158\,\Delta\,S. 
\end{equation}

Taking the value of ${\rm [Fe/H]}=-0.53$ (Armandroff \& Zinn 1988) and 
calculating the $\Delta\,S$ value, we find that the Suntzeff et al. (1991)
calibration gives colors which are $\sim 0.02$~mag bluer than those 
given by the Blanco (1992) calibration.  Reddenings derived from 
Clementini et al.'s (1995) recalibration of the delta-S index are within 
0.01 mag of those obtained with the Blanco calibration, being slightly 
larger.  Table~9 gives our reddening determinations for RRab 
variables with good light curves using the Blanco calibration.  Some 
variables had not yet achieved minimum light in the 0.5-0.8 phase range 
after maximum light.  The points in this range were averaged to find 
$\langle{\bv}\rangle$ in all cases.  The reddenings found in this
way for the stars labeled as ``bright and red" may be incorrect, since
those stars may in fact be unresolved blended images, as noted above.  
The mean reddening
value for the remaining 10 RRab stars which are believed to be probable
members of NGC~6441 is $E(\bv) = 0.51 \pm 0.02$~mag.  The range in
reddening values is consistent with previous determinations (see
Layden et al. 1999) that the NGC~6441 field is subject to significant
differential reddening.   

Also shown in Table~9 is a comparison to the reddening
values found by Layden et al. (1999). We see that our reddening 
determinations are generally higher than those found by Layden et al.  The mean 
reddening value determined from the six normal RRab stars observed by 
Layden et al. is $E(\bv) = 0.45 \pm 0.02$~mag. It is uncertain to what 
extent this difference arises from the discrepancy in the $V$ magnitudes 
between our data and theirs since we have no way to compare with their 
$I$ data in order to check their calibration of 
$E(V-I)$ to $E(\bv)$.  Our mean reddening value for NGC~6441 can also 
be compared to the result of Hesser \& Hartwick (1976), who determined 
$E(\bv)=0.46\pm0.15$~mag, and the results of Zinn (1980) and Reed et al. 
(1988), who obtained $E(\bv)=0.47$ and 0.49~mag, respectively, from their 
analyses of the integrated cluster light. 

Adopting $A_V = 3.2~E(\bv) = 1.63$ and a range of +0.8 to +0.5 for $M_{\rm V}$, we find 
the distance modulus for NGC 6441 in the range 15.08--15.38, given 
$V_{\rm RR}=17.51$.  The distance to the cluster is estimated to be from 10.4 to 
11.9 kpc.  In comparison, Layden et al. (1999) estimated the distance to 
the cluster to be 11 to 13 kpc.  Harris (1996) lists NGC 6441 as having a 
distance of 11.2 kpc with $E(\bv) = 0.44$.

Heitsch \& Richtler (1999), in an analysis of NGC~6441, found the total 
reddening to be $E(V-I)=0.49 \pm 0.03$ with a maximum differential reddening of 
$\Delta [E(V-I)] = 0.20$.  Layden et al. (1999) found $E(V-I) = 0.577$ based on the 
average RRL reddening.  This difference may be again attributable to the zero 
point offset we found in the $V$ data of Layden et al. (see \S2).  In 
comparing the $V$,$I$ color-magnitude diagram of Layden et al. to Heitsch \& 
Richtler, we see that the HB is offset to a brighter magnitude for 
Layden et al.  Heitsch \& Richtler found $V_{\rm HB} = 17.51 \pm 0.07$, which is 
also in good agreement with our determination from the RRL.  With these 
data they determined the distance to NGC 6441 to be $13.8 \pm 1.3$ kpc.  
The difference between our distance determination and theirs can be found in 
the adopted metallicity for NGC 6441 and in the reddening which in turn 
affects the adopted extinction.

One should note that the ab-type RRL in NGC~6441 by their very nature are 
different from those which Blanco used in establishing his relationship
between metallicity, period, and intrinsic color.  We have assumed here
that his formula is applicable to the RRab stars in NGC~6441.  This
might not be the case.  Bono et al. (1997) have argued on theoretical
grounds that the red edge to the instability strip might lie at
lower effective temperatures for metal-rich than for metal-poor RRLs
of equal luminosity.  If so, and if, as argued in this paper, the RRab 
stars in NGC~6441 are unusually bright for their metallicity,
then Blanco's reddening calibration
might not apply perfectly to the RRab stars in NGC~6441.

We would also like to point out that we have assumed that the RRL in NGC~
6441 are metal-rich in our calculation of the reddening.  If the RRL are 
metal-poor (see \S5.5), the derived reddening would be even higher.  If one 
assumes ${\rm [Fe/H]} = -2.0$ for the RRL, the derived reddening increases 
by 0.05 mag, which would be much redder than the other derived reddenings 
for NGC 6441. 

We note that the {\sc cobe/dirbe} dust maps of Schlegel, Finkbeiner, \& 
Davis (1998) imply a high reddening $E(\bv) \simeq 0.66$~mag for NGC~6441.  
There is evidence suggesting that the Schlegel et al. maps do overestimate 
the reddening for highly reddened regions (Arce \& Goodman 1999; 
von Braun \& Mateo 2001).  Our reddening value for NGC~6441 clearly supports 
this conclusion.

\subsection{Eclipsing Binaries and LPVs} 

We were able to find a number of eclipsing
binary stars within our field of view, which are not likely to be members 
of NGC~6441.  The 
binaries listed by Layden et al. (1999) were all recovered.  Table~10 lists 
photometric data for the binary stars.  Due to our sampling it was 
somewhat difficult to determine accurate periods for detached binaries.  

Our observations were not geared toward locating LPVs, 
but we were able to detect some stars exhibiting luminosity changes over 
our 10 day run.  These stars and their locations are listed in Table~2.  We 
were able to reidentify a small number of LPVs already found by Layden et al. 
(1999) (V1, V2, V5, V6, V9, and V10) along with a couple of probable new LPVs 
(V98 and V99).

\subsection{Suspected Variable Stars} 

Following the naming scheme by Layden et al. (1999), we list in Table~11 
stars which exhibit variability, but were difficult to classify for reasons 
such as high scatter in the light curves or abnormally low amplitudes.   
SV10, SV11, and SV12 all show some structure to their light curves giving an
indication they may be short period variables (see Figure 5).  Yet, 
the scatter in the curves matched with their low amplitudes, ranging 
from 0.10 to 0.15~mag in $B$ and $V$, make them difficult to classify.  
SV13 appears to vary in magnitude over a period of days yet the amplitude 
found is only 0.15~mag.  The photometry of both SV14 and SV15 shows that 
the magnitudes of the stars have an amplitude of about 0.4~mag.  Again, 
scatter in the curves makes the classification of these stars unknown.

\section{Discussion}

\subsection{General Properties of NGC~6441 RR Lyrae} 

The higher number of RRL now found in NGC~6441 gives us the opportunity 
to better analyze the properties of this cluster.  In Table~12, we list 
the average properties of the RRL found in NGC~6441 and compare it 
to the clusters M3 and M15, typical Oosterhoff I and II clusters.  As was 
shown in Layden et al. (1999) and Pritzl et al. (2000), the 
RRL in NGC~6441 have unusually long periods for a metal-rich globular
cluster.   Clusters as metal-rich as NGC~6441 usually have few if any
RRL stars.  If indeed such a cluster did have numerous RRLs, then, from
what is known of metal-rich field RRLs, one would expect the mean period
of its RRab stars to be even shorter than those of Oosterhoff type I
globular clusters.  In actuality, the long mean periods of the RRL in 
NGC~6441, and 
its high $N_{\rm c}/N_{\rm RR}$ ratio (where $N_{\rm c}$ is the number 
of RRc stars and $N_{\rm RR}$
is the total number of RRL in the system), are closer to the values
expected in a metal-poor Oosterhoff II globular cluster. The value of 
$\langle{P_{\rm ab}}\rangle$ for the RRL is long even for Oosterhoff II 
systems.     

In Figure 8, we plot the mean periods of the RRab stars in NGC~6441 and 
in Oosterhoff I and Oosterhoff II clusters as a function of their parent
cluster metallicity, including NGC~6388 (Pritzl et al. 2001).  As was shown 
in Pritzl et al. (2000), 
we see how the trend of decreasing period with increasing metallicity for 
Galactic globular clusters is broken by NGC~6441.  We conclude,
as did Pritzl et al. (2000), that NGC~6441 does not fall into either of the 
usual Oosterhoff groups in this diagram.  

It has also been argued that evolutionary effects may be the cause of 
the Oosterhoff dichotomy.  Clement \& Shelton (1999), Lee \& Carney (1999), 
and Clement \& Rowe (2000) have argued that the location of RRab stars in the 
period-amplitude diagram is more a function of Oosterhoff type rather than 
metal abundance.  Indeed, the RRab stars in Figure 7 do fall along the 
line for RRab stars in Oosterhoff II clusters as given by Clement (2000; 
private communication). 

We performed the compatibility condition test of Jurcsik \& 
Kov\'{a}cs (1996) on the RRab stars in NGC~6441.  The $D_m$ values are listed in 
Table~6.  We find that only 4 of 14 RRab stars have a value of $D_m$ 
smaller than the cutoff of 5.0 suggested by Clement \& Shelton (1997) as 
appropriate for identifying ``normal" RRab light curves from Fourier fits of 
the order we used.  The interpretation of this result in the case of the 
NGC~6441 variables is uncertain, however.  Large values of $D_m$ can be 
caused by light curves which do not have ``normal" shapes, but can also 
result from the analysis of low quality light curves.  In addition, the 
applicability of the $D_m$ criterion to the characterization of the very long 
period RRab light curves found in NGC~6441 has not been established.  Because 
of the short time span covered by our observations, it is also possible that 
stars showing the Blazhko effect might not be identified as such by 
inspection or by the $D_m$ criterion in our dataset.  Due to these 
uncertainties, in the following analysis we will include NGC~6441 variables 
which have light curves showing little scatter and having no significant 
gaps, rather than include only variables with values of $D_m$ smaller than 
5.0.

Clement \& Rowe (2000) also used the masses and luminosities of the RRc 
stars to illustrate the differences between the Oosterhoff groups.  
Their Table~4 shows 7 globular clusters 
arranged in order of increasing $\log\,(L/L_{\odot})$ for the RRc variables, 
where the RRc luminosities have been determined from Fourier parameter 
$\phi_{31}$, where $\phi_{31} = \phi_{3} - 3 \phi_{1}$.  
Making use of their Eqs. 2-5, which derive from Simon \& 
Clement (1993a, 1993b), and 
the Fourier parameters listed in  Table~6, we have calculated the masses, 
luminosities, temperatures, and absolute magnitudes for a small number of 
RRc stars in NGC~6441 (see Table~13).  Following the criteria for acceptable 
stars in their paper 
we excluded all long period RRc stars (see \S5.2) since these stars 
were shown to have abnormally low derived masses and included only 
the RRc stars with an error in $\phi_{\rm 31}$ below a certain limit.  
Our cutoff for the error in the $\phi_{\rm 31}$ 
measurement was 0.50, instead of 0.20 used by Clement \& Rowe, to allow a 
larger sampling 
of stars.  After this restriction we only kept RRc stars with light curves 
having low scatter and low errors for the data points.  The RRc with the 
best light curve was V70. 
Leaving out V71 and V77, for which the Simon \& Clement equations yield 
especially low masses, we find the mean values for the RRc star mass, 
$\log\,(L/L_{\odot})$, $T_{\rm eff}$, and $M_V$ are 
$0.47\,M_{\odot}$, 1.65, 7408~K, and 0.79.  If we were to include V71 and 
V77, the mean values would change to $0.44\,M_{\odot}$, 1.65, 7403~K, and 0.79, 
which are not significant changes.  
By the criterion of either mean luminosity or mean $T_{\rm eff}$,
the RRc stars of NGC~6441 appear to fall among the Oosterhoff I clusters in
Clement \& Rowe's Table~4.  This, of course, contradicts the result
obtained above from the RRab period-amplitude diagram.

It should be noted that the masses calculated for {\it all} of the RRc stars 
tend to be low.  The mean mass of $0.47\,M_{\odot}$ is about $0.02\,M_{\odot}$
smaller than the canonical helium core mass at the helium flash (cf. Table~1 
in Sweigart 1987; see also Sweigart 1994 and Catelan, de Freitas Pachaco, \& 
Horvath 1996 for extensive discussions of possible sources of uncertainty 
in the helium core mass at the helium flash).  It is unlikely that 
a star with this mass would become a RRL star.  Scatter in the light curves 
and/or a lack of highly sampled light 
curves may lead to uncertainties in the Fourier decomposition.  The sinusoidal 
shape of the light curves of the RRc stars results in large uncertainties in 
the $\phi_{\rm 21}$ and $\phi_{\rm 31}$ estimates due to the $\phi_{\rm 2}$ 
and $\phi_{\rm 3}$ phases being determined with low precision.  While this 
may lead to scatter in the derived RRc parameters, it cannot account for the 
systematically low values for the masses.  One possible reason for the low 
mass values could be that a problem with the zero point in the Simon \& Clement (1993a, 
1993b) 
calibration for the RRc masses exists.  In any case, it appears that even 
in this 
Oosterhoff classification based on derived physical parameters for the RRc stars, 
NGC~6441 has difficulty being classified as either an Oosterhoff type I or 
Oosterhoff type II cluster.  
(Further points on this classification 
scheme are discussed in \S5.4, \S5.5, and \S6.)  Another possible 
explanation for the low RRc masses is that the Fourier relations derived by 
Simon \& Clement (1993a, 1993b) may not be applicable to the RRc in NGC 6441, 
especially if some noncanonical effect is responsible for the blue extension 
of the HB.

Given the existence of long period RRc stars in NGC 6441 (\S5.2), and the 
unusually low masses obtained for the RRc stars with periods near 0.3 days, 
one might wonder if some of the shorter period RRL might be second overtone 
pulsators.  However, the light curves for these stars appear not very 
different from those of shorter period RRc stars in other clusters, and the 
location of these stars in the $A_{\rm 21}$ versus $\phi_{\rm 21}$ diagram 
(Figure 6) is also broadly consistent with the locations of RRc stars in other 
clusters.  It was also shown by Clement \& Rowe (2000) that potential 
second-overtone RRL tend to have $V$ amplitudes less than 0.3 mag.  This 
is not the case for the RRc stars in NGC~6441.

Parameters for the RRab stars in NGC 6441 were derived using 
the Jurcsik-Kov\'{a}cs method (Jurcsik \& Kov\'{a}cs 1996; Kov\'{a}cs \& 
Jurcsik 1997).  
The Fourier parameters for the RRab stars with good light curves are listed in 
Table~6 and were corrected to work in the Jurcsik-Kov\'{a}cs method which 
is based on a sine decomposition.  Equations 1, 2, 5, 11, 17, and 22 from 
Jurcsik (1998) were used to 
calculate the parameters listed in Table~14.  The values of 
$\log\,(L/L_{\odot})$ and $\log\,T_{\rm eff}$ were increased by 0.1 and 0.016, 
respectively, following the prescription in Jurcsik \& Kov\'{a}cs (1999).  
The mean values for the mass, 
$\log\,(L/L_{\odot})$, $\log\,T_{\rm eff}$, $M_{\rm V}$, and [Fe/H] are 
0.54~$M_{\odot}\pm0.01$, $1.66\pm0.02$, $3.82\pm0.01$, $0.68\pm0.03$, and 
$-0.99\pm0.06$, respectively.  Comparing these results to the empirical data in 
Figure 1 of Jurcsik \& Kov\'{a}cs (1999), we see that the mean value of 
$\log\,(L/L_{\odot})$ for NGC 6441 is about 0.05 brighter than the data from 
Jurcsik \& Kov\'{a}cs at the 
metallicity of -0.53.  This agrees with our assertion that the RRL in NGC 6441 
are unusually bright for their metallicity.  While the value for the mean 
mass of the NGC 6441 RRab stars is consistent with the cluster metallicity, 
the mean value for $\log\,T_{\rm eff}$ is about 0.02 lower than the data given 
by Jurcsik \& Kov\'{a}cs (1999).  Comparing the mean value of $M_V$ given 
by the RRab parameters against the value given by Eq. 5 in Kov\'{a}cs \& 
Jurcsik (1996) relating $M_V$ to [Fe/H], the absolute magnitude given 
by our RRab stars is about 0.25 mag brighter, where $M_V = 0.19 
{\rm [Fe/H]} + 1.04 = 0.94$ for ${\rm [Fe/H]} = -0.53$.

Possibly the most interesting result to come out of the analysis of 
the RRab is the low metallicity of ${\rm [Fe/H]} = -0.99$.  According to Eq. 4 of 
Jurcsik (1995), this metallicity is actually -1.3 on the Zinn \& West (1984) 
scale.  This seems unusually low when compared to the cluster metallicity 
derived by Armandroff \& Zinn (1988) of -0.53.  There is some question as 
to the validity of applying the Jurcsik-Kov\'{a}cs method to the RRab stars in 
NGC~6441, given their unusually longer periods.  We nonetheless note that the 
mean metallicity we derive in this way is close to that predicted by the HB 
models of Sweigart (2001), which yeilded a best fitting track for a [Fe/H] 
of approximately -1.4, assuming an $\alpha$ enhancement of 
${\rm [\alpha/Fe]} = +0.3$.  Such a low metallicity for the RRab stars would 
of course imply that there is a real spread in metallicity among the 
NGC~6441 stars (\S5.5).

As pointed out by Pritzl et al. (2000), NGC~6441 contains an unusually large 
number of long period ($P > 0.8$ day) RRab.  Considering all candidate RRab 
stars believed to be members of NGC~6441, we find $38\%$ (10 out of 26) of 
the RRab to have periods greater than 0.8 day.  Although other clusters 
are known to contain long period RRL such as these, most notably $\omega$ 
Centauri, to our knowledge no other cluster has such a large proportion of 
long period RRab.  This large proportion of very long period RRab stars, 
a property NGC~6388 shares with NGC~6441 (Pritzl et al. 2000, 2001), is thus 
an additional way in which NGC~6441 is distinguished from typical 
Oosterhoff type II clusters.

\subsection{RRc Variables} 

Kemper (1982) showed that there are few metal-rich RRc stars in the solar 
neighborhood. RRLs of any type are rare in the more metal-rich globular
clusters.
The unusual nature of NGC~6441 gives us an opportunity to investigate c-type 
RRL in an environment more metal rich than those in which they
are usually found, either in globular clusters 
or in the field.   
Although the periods of a few of the RRc in NGC~6441 do tend to fall at longer 
values, we see that $\langle{P_{\rm c}}\rangle$, as given in Table~12, 
is not unusually large 
compared to values found in Oosterhoff II globular clusters (Sandage 1982).  

The light curves of the NGC~6441 RRc stars also seem to have some 
distinguishing features.  As the period goes to longer values, the bump 
seen during rising brightness tends to be found at earlier phases.  For most, 
but not all, of the 
shorter period RRc we find the bump occurring at a phase $\sim 0.2$ before 
maximum while 
for the longer period ones, such as V69 and V76, it occurs at $\sim 0.3$ 
before maximum.  Recent progress in the nonlinear, convective hydrodynamic 
modeling of the light curves of RRc stars (Bono, Castellani, \& Marconi 2000) 
may soon determine the physical properties responsible for such peculiarities. 

Layden et al. (1999) mentioned that the light curves  of NGC~6441 RRc stars exhibit 
longer than usual rise times.  They comment that the RRc stars of NGC~6441 
have a phase interval of ``rising light" between minimum and maximum 
brightness greater than $\sim 0.5$.  While the longer period RRc variables do 
have rise intervals around 0.5, we find that on average, most were 
$\sim 0.42-0.45$.  This is in the higher end of the range listed by Layden et al. 
for RRc variables from the \textit{General Catalog of Variable Stars} 
(Kholopov et al. 1985).  There seems to be a slight trend of increasing rise time with 
increasing period.  

Layden et al. (1999) also noted that the minima for the RRc stars may 
be uncharacteristically sharp, pointing to SV3 (V70).  We find 
that SV3 is better classified as an eclipsing binary star.  
We do not find any unusual sharpness to the minima of the RRc variables. 

The long period of V69 results in an unusually short gap between
the period of the longest period RRc star and the period of the
shortest period RRab star in NGC~6441.  If the RRab and RRc stars
had the same mass and luminosity, and were there a single
transition line in effective temperature which divided RRab from
RRc pulsators, then we would expect a gap of about 0.12 between the
logarithms of the periods of the longest period RRc star and the shortest
period RRab star (van Albada \& Baker 1973).  Clearly, we do not
see that, indicating that one of those assumptions may be in error.
Again, however, the existence of differential reddening in the field
makes it difficult to interpret the CMD at the level which
one would like in addressing this point.

With a better understanding as to which stars are RRc variables, we update 
the histogram over period for NGC~6441 RRLs in Figure 9.  As noted in 
Pritzl et al. (2000), the distribution of RRc to RRab stars in NGC~6441 
shows that this cluster is relatively rich in RRc stars which 
is similar to other Oosterhoff II clusters, contradicting what one would 
expect from its metallicity.  Again, we see how NGC 6441 stands out as 
anomalous when compared to other Oosterhoff clusters.

\subsection{Period-Amplitude Diagram} 

The period-amplitude diagram provides a way to look at the general 
trends of the RRL in a system without having to worry about reddening.  
In Figure 10 
we revisit the diagram presented in Pritzl et al. (2000), comparing 
NGC~6441 to other 
globular clusters and to field stars of similar metallicity (M15: Silbermann 
\& Smith 1995, Bingham et al. 1994; M68: Walker 1994; M3:  Carretta et al. 
1998; 47 Tuc: Carney et al. 1993).  We see, 
compared to the 
metal-rich field stars, the RRL of NGC~6441 fall at unusually long 
periods.  They even stand out compared to the RRL in the Oosterhoff Type II 
globular clusters M15 and M68.  As was noted by Pritzl et al.,  
the trend of the increasing period with decreasing metallicity for a given 
amplitude is broken by this cluster.  It is interesting to note that the 
single RRab star known to be a member of 47 Tuc, V9, 
falls among the RRL found in NGC~6441. We also note that we have assumed 
in this
discussion that the metal-abundance of the RRL stars in NGC~6441 is the
same as that found by Armandroff \& Zinn (1988) for the cluster as a whole.
As of yet, we have no direct measurement
of metallicity for individual RRL stars (see \S5.5). 

The assertion that the RRab in NGC 6441 stand out when compared to M68 
and M15 seems to contradict our previous finding that the RRab of 
NGC 6441 fall along the line for RRab in Oosterhoff II clusters as given by 
Clement (2000; private communication) (see \S5.1).  One key to the 
analysis done by Clement (2000), Clement \& Rowe (2000), and Clement \& 
Shelton (1999) is to determine which variables have ``normal" light 
curves.  This was determined by the Jurcsik \& Kov\'{a}cs (1996) 
compatibility condition.  For example, according to this requirement, 
Clement \& Shelton showed that only V23 and V35 in M68 had ``normal" 
light curves.  These two stars do fall along the Oosterhoff II line 
implying that the RRab in NGC 6441 do not stand apart when considering 
those stars with ``normal" light curves.  The case of M15 is not as 
easily resolved.  When plotting the Oosterhoff I and II lines as 
given by Clement against the M15 RRab data, a few RRab stars fall near 
each line with a majority of them lying in-between the two lines.  
We have applied the compatibility test on the M15 RRab variables with 
the best light curves.  Table~15 lists the data.  Fourier parameters for the 
Bingham et al. (1984) data were taken from Kov\'{a}cs, Shlosman, \& Buchler 
(1986).  We performed the Fourier analysis on the RRab variables from 
Silbermann \& Smith (1995).  While a couple of M15 RRab with ``normal" light 
curves are found near either Oosterhoff line, there are a few that are 
found between the lines.  We do note that the Bingham et al. data for M15 
is photographic which may lead to some discrepancies in the amplitudes.  There 
are also some possible uncertainties in the compatibility condition (see
\S5.1). 

The period shift of the RRab stars in NGC~6441 relative to RRab stars
of equal $B$ amplitude in the globular cluster M3 (from Sandage et al.
1981) is about 0.08 in $\log\, P$.  If the masses of the RRab stars in M3
and NGC~6441 were the same, from Eq. 2 of van Albada \& Baker (1971) 
this would correspond to a difference of
$\Delta \log\,L$ = 0.10 or $\Delta\,M_{\rm bol} = 0.24$.  This, admittedly 
simplified, analysis would make the RRab stars in NGC~6441 slightly more 
luminous than the RRab stars within the Oosterhoff II cluster M15.

For the luminosity difference estimated in this way to become zero as a
consequence of a possible mass difference, the NGC~6441 RRL masses
would have to be smaller by $\Delta\,\log\,M \approx 0.12$.  Taking
the M3 RRL masses to be $\langle M_{\rm RR}\rangle \approx 0.64\,
M_{\sun}$ (as estimated from Eq. 7 of Sandage 1993, using
${\rm [Fe/H]} = -1.55$ for the cluster metallicity), this would
imply $\langle M_{\rm RR}\rangle \approx 0.48\, M_{\sun}$ for
NGC~6441 -- which is unreasonably small, being smaller than the
helium core mass at the helium flash.  Yet it is similar to that 
found by Fourier analysis (see \S5.1).  

On the other hand, the Sweigart \& Catelan (1998) synthetic HB's 
assuming $Y_{\rm MS} = 0.23$, $Z = 0.006$ for both
the canonical and the helium-mixing scenarios [the latter with
the Reimers (1975) mass-loss parameter $\eta = 0.6$] imply mean masses of 
$\langle M_{\rm RR}\rangle \approx 0.58\, M_{\sun}$ for the NGC 6441 RRab 
variables.  For comparison a canonical 
synthetic HB for an M3-like HB morphology with $Y_{\rm MS} = 0.23$ and 
$Z = 0.001$ was computed following the same methods as in Sweigart \& 
Catelan and using the HB tracks from Catelan et al. (1998), giving 
$\langle M_{\rm RR} \rangle \approx 0.65\, M_{\odot}$.  Taking this 
theoretical mass difference into account, one obtains a revised luminosity 
difference of $\Delta\,M_{\rm bol} \simeq 0.14$~mag between the RRab 
stars in M3 and NGC~6441, with those in NGC~6441 still being the brighter.

\subsection{Comparisons to Long Period RR Lyrae in $\omega$ Centauri} 

Overall, the RRL stars in $\omega$ Centauri are similar in period and 
amplitude to those which would be found within a typical Oosterhoff type II 
system.  Butler et al. (1978), Caputo (1981), and Clement \& Rowe (2000) have 
noted, however, that $\omega$ Centauri might contain a few RRL similar to those 
in an Oosterhoff type I cluster.  In addition, as noted in Pritzl et al. (2000), 
$\omega$ Centauri contains a number of very long period RRab stars, 
similar in period and amplitude to the longer period RRab stars in NGC~6441. 

Here we note that $\omega$ Cen also contains a number of RRc variables of 
unusually long period, similar to V69 and V76 in NGC~6441.  Making use of the 
data from Petersen (1994), we plot the Fourier parameters $\phi_{21}$ vs. 
$A_{21}$ for $\omega$ Cen (Figure 11).  A noticeable trend is exhibited here.  
The longer period RRc variables lie as a distinct group at shorter $\phi_{21}$ 
($\phi_{21} \approx 2$).  
A similar trend towards shorter $\phi_{21}$ is noticeable in Figure 6 for 
NGC~6441.  In Figure 12, we plot some of the RRc stars 
in $\omega$ Cen in a period-amplitude diagram.  The periods and amplitudes 
were taken from Kaluzny et al. (1997).  When there was more than one entry 
for a single star, the values were averaged.  The [Fe/H] values come from 
Rey et al. (2000) and the RRc classifications were taken from Butler et al. 
(1978).  We see that although there seems to be a trend of 
increasing amplitude, decreasing period with increasing metallicity, there are 
some longer period RRc 
found in the more intermediate metallicity range.  

An attempt was made to search through the literature for other long-period 
c-type RRL in other globular clusters.  One clear case was found, 
V70 in M3, with 
$P=0.49$~d and $A_V = 0.35$~mag, which was misclassified by Kaluzny et al. (1998) 
as an RRab variable, but correctly identified as an RRc by Carretta et al. 
(1998).  Another possible long-period RRc star is V76 in M5 (Kaluzny et al. 
2000).  Its period of 0.432544~d makes it unusually long as compared to the 
other RRc stars in M5.  The lack of such stars in other clusters once again 
demonstrates the unusual nature of NGC~6441 and NGC~6388, and provides 
another way in which the clusters appear to be different from typical 
Oosterhoff type II systems.  

This suggests that the RRL variables in NGC~6441 may, perhaps
surprisingly, belong to a metal-poor component similar to the one
to which the long-period RRc's in $\omega$~Cen belong (see Figure 12).  
Again, we note that 
there are no direct metallicity measurements for the RRL in NGC~6441, and that, 
in this paper, we have hitherto assumed them to have the overall cluster 
[Fe/H] value.

\subsection{A Metallicity Spread in NGC~6441?} 

Piotto et al. (1997) first noted that the RGB of NGC~6441 shows  
an intrinsic spread in color. After rejecting the possibility 
that differential reddening is the cause of the elongated, 
sloping HB, they remarked that ``at least qualitatively, a spread 
in metallicity might be an appealing explanation for the anomalous 
HB and for the spread in the RGB." The RRc stars in the cluster 
do appear to give some support to their hypothesis. While detailed 
calculations are needed to verify this possibility (Sweigart 2001), 
we also note that: 

i) Given the NGC~6441 HB number counts (Zoccali 1999, private 
communication), the RRL and blue HB components comprise only  
$\sim 15\%$ of the total HB population of NGC~6441. Therefore, if 
these HB components are the progeny of metal-poor RGB stars in  
the cluster, only a similar small fraction of metal-poor RGB stars should be 
present in NGC~6441 (and similarly in NGC~6388). Such a metal-poor 
component could presumably be fairly easily ``hidden" behind the 
differential reddening that is present in the cluster (e.g., 
Layden et al. 1999); 

ii) Since the number ratio of RRc to RRab stars in NGC~6441 is 
more typical of Oosterhoff type II than Oosterhoff type I clusters, 
one might think that the 
usual ``evolutionary mechanism" invoked to account for the c/ab
number ratios (van Albada \& Baker 1973) could be at play in this 
cluster as well, implying that the NGC~6441 RRL variables evolve 
from blue to red in the CMD. Redward evolution for stars crossing 
the instability strip is thought to be a likely characteristic of 
metal-poor globular clusters with primarily blue HB's (e.g., Lee, 
Demarque, \& Zinn 1990), which  
might also imply that the blue HB plus RRL component in NGC~6441 is 
metal-poor; 

iii) Clement (2000), adopting a strict selection criterion to include 
only RRab variables with ``regular" light curves (Jurcsik \& Kov\'{a}cs 
1996), found that the NGC~6441 RRab stars occupy a location in the 
Oosterhoff type II systems (see also Fig.~7).  Pritzl et al. (2000), 
and our updated 
Figure 10, show the RRab stars in NGC~6441 to be at least as long 
in period at a given amplitude as the RRab stars in typical Oosterhoff 
type II systems.  That might suggest that the NGC~6441 RRab stars are 
metal-poor, given that the most metal-rich Oosterhoff type II systems 
known have ${\rm [Fe/H]} \simeq -1.6$ (Ortolani et al. 2000);

iv) Moehler, Sweigart, \& Catelan (1999) have shown that the 
gravities of the blue-HB stars in NGC~6441 (and NGC~6388) are not 
consistent with the predictions based on the non-canonical scenarios
proposed by Sweigart \& Catelan (1998). On the other hand, their 
measured gravities are not clearly inconsistent with canonical, 
metal-poor models. 

Hence it appears that the Piotto et al. (1997) conjecture that a
spread in metallicity would be qualitatively consistent with the 
anomalous HB of NGC~6441 is supported by some of the currently available 
data for the RRL variables and blue-HB stars in the cluster. But 
can a spread in metallicity lead to a sloping HB as seen in the 
cluster as well? 

Apparently, yes. If we assume, as is currently believed, that the 
mean slope of the $M_V({\rm HB}) - {\rm [Fe/H]}$ relation is in 
the range $\sim 0.2-0.3$, and if the more metal-rich component 
(i.e., the bulk) of the cluster has 
${\rm [Fe/H]} \approx -0.5$~dex, this would imply that the more 
metal-poor component of the cluster would have a metallicity of 
the order ${\rm [Fe/H]} \approx -2.2 \rightarrow -3.0$~dex. While 
this may seem somewhat too low, there are several effects which 
may, at least in principle, contribute to the production of an HB 
slope and hence alleviate somewhat the constraints on the 
metallicity of the metal-poor component, such as: a)~An increase 
in the slope of the $M_V({\rm HB}) - {\rm [Fe/H]}$ relation for 
${\rm [Fe/H]} > -1$ (e.g., Castellani, Chieffi, \& Pulone 1991); 
b)~Helium mixing and/or an increase in the core mass at the 
helium flash for the more metal-poor components (Sweigart 1997); 
c)~Evolution away from a blue,
metal-poor zero-age HB, which might make any evolved metal-poor 
RRL stars brighter than the ``mean" $M_V({\rm HB}) - {\rm [Fe/H]}$ 
relation (e.g., Lee \& Carney 1999); d)~A ``canonical slope" of the 
zero-age HB (Brocato et al. 1999), which, though small 
($\Delta\,V \simeq 0.1$~mag), would contribute to producing a 
sloping HB; e)~Differential reddening, which is clearly present 
in the field of the cluster (\S4.5; Layden et al. 1999). 
 
These arguments are obviously of a qualitative nature.  However, 
Sweigart (2001) has recently carried out stellar evolution calculations 
to explore this metallicity-spread scenario in more detail.  
These evolutionary models show that a spread in metallicty can 
naturally lead to an upward sloping HB under the simple assumptions 
that all of the stars are coeval and that the mass loss efficiency, 
as measured by the Reimers (1975) mass loss parameter, is independent 
of [Fe/H].  Moreover the size of the predicted HB slope is close 
to that observed in NGC~6441.  

However, among the issues that still need to be addressed, we highlight 
the following possible caveats: 
 
i) Why are NGC~6441 and $\omega$~Cen able to produce long-period 
RRc variables, which seem to be extremely rare in single-metallicity 
globular clusters (see \S5.4)? Can this be somehow related to the extreme deep 
mixing signatures which have been identified in $\omega$~Cen 
(Norris \& Da Costa 1995a, 1995b)? Obviously, abundance measurements 
are badly needed for both the red giants and HB stars in NGC~6441 
(and NGC~6388); 
 
ii) Is the HB luminosity function of NGC~6441 consistent with a 
sloping HB caused by a metallicity sequence, with 
mean metallicity decreasing with bluer colors? Since at any given 
metallicity globular cluster HBs show a large spread in color, 
a superposition of ``template" HBs of globular clusters 
of different metallicity (e.g., 47~Tuc at the metal-rich end, 
NGC~362, M3, M2, M15, all the way down to NGC~5053 at the most 
metal-poor end) would certainly not lead to a ``tight" sloping 
HB, but instead to a large color/luminosity scatter at any 
given luminosity/color; 
 
iii) Following the same reasoning as above, the RRL population 
might also be expected to have contributions from several 
different metallicity values. Would this be consistent with the 
overall brightness of the NGC~6441 RRL? How would this be 
reconciled with the fairly tight period-amplitude distribution 
of the cluster stars?  

iv) What is the connection between the bright NGC~6441 
RRL stars and V9 in 47~Tuc?  While a correlation between position 
on the HB and mixing signatures has been clearly demonstrated in 
the case of 47~Tuc (Briley 1997), a spread in metallicity appears 
rather unlikely for this cluster (Suntzeff 1993).  As a matter of fact, 
V9 itself 
has been conclusively shown to be only moderately metal-deficient 
(Keith \& Butler 1980; Smith 1984), with a metal abundance consistent 
with that of 47~Tuc;

v) If the RRL in NGC~6441 are indeed very metal poor, one would expect 
them to be more massive than those in Oosterhoff type I clusters 
(e.g., Sandage 1993 and references therein).  However, our derived RRc 
masses (\S5.1) are very low -- even lower than those derived using 
the same method for Oosterhoff I globular clusters such as M3, M5, 
NGC~6171 (cf. Table~3 in Kaluzny et al. 2000, and Table~4 in 
Clement \& Rowe 2000) and NGC~6229 (Borissova, Catelan, \& Valchev 
2001). Can this trend be reconciled with the ``metal-poor" scenario?
On the other hand, recent work by Sohn et al. (2001) has shown that 
the RRc masses derived using this method appear to be very low for the 
${\rm [Fe/H]} \simeq -2$, Oosterhoff II cluster M53 (NGC~5024).

\section{Classification}

It has been shown throughout this paper, along with the discussion in 
Pritzl et al. (2000), that the classification of NGC 6441 into one or
the other of the usual Oosterhoff groups is difficult. The long 
periods of the RRab stars and the relatively large proportion of c type
pulsators are more typical of Oosterhoff II than Oosterhoff I systems, though we
again note that $\langle P_{\rm ab} \rangle$ is unusually large even for 
Oosterhoff II. In
contrast, the mean luminosity of the NGC 6441 RRc stars as found 
by applying the Simon \& Clement (1993a, 1993b) method (see \S5.1) is 
consistent 
with values found in Oosterhoff I clusters such as M3.  This is in conflict with 
our assertion that the RRL are unusually bright (see \S5.3).  Figure 7 shows 
that the NGC~6441 RRab 
stars do appear to fall along the line in the period-amplitude diagram 
for RRab stars in Oosterhoff II clusters, as given by Clement \& Shelton (1999).  
However, using a somewhat different sample of comparison stars, we find 
in Figure~10 that the NGC~6441 stars tend to have slightly longer 
periods at a given amplitude than the bulk of the RRab stars in 
the Oosterhoff II clusters M15 and M68.

The mere fact that NGC 6441 has a blue extension to its HB that extends 
through the instability strip makes it unusual when compared to other 
metal-rich globular clusters.  It may be the case that RRL that form in 
metal-rich globular clusters exhibit different properties than metal-rich 
stars that form in the field.  As shown in Figure 10, V9 in 47 Tuc falls in 
the same region of the period-amplitude diagram as the RRab in NGC~6441 and 
NGC~6388.  Recently, a RRab star was found in the metal-rich 
cluster Terzan 5 (Edmonds et al. 2001).  While its period ($\sim$0.61 d) is shorter 
than the average periods of the RRab in NGC 6441 and NGC 6388 and V9 in 47 Tuc, 
it is longer than field stars of similar metallicities.  These examples (see 
also Layden 1995, esp. his Fig.~1) all 
illustrate that these clusters may be exceptions to the Oosterhoff 
classification scheme and to the usual correlation of RRL properties
with metallicity among field RRL.

\section{Summary and Conclusions}

NGC~6441 stands out as one of the more unique globular clusters 
of our Galaxy.  NGC~6441 is confirmed to be a metal-rich 
globular cluster exhibiting an unusual horizontal branch morphology.  A strong 
red component of the horizontal branch is seen, as is expected for a cluster 
in this metallicity range.  In addition to the red clump, the cluster has a 
blue horizontal branch extending through the instability strip which is 
not found in other 
clusters of similar metallicities except for NGC~6388, i.e., NGC~6441 exhibits 
a second-parameter effect.  The explanations of such an effect may 
be constrained by the sloped nature of the horizontal branch, getting brighter 
in $V$ with decreasing $\bv$, as Sweigart \& Catelan (1998) suggested.  
It is also possible that NGC~6441 may have a spread in its metallicity 
making it analogous to $\omega$ Centauri, in which case its classification as a 
``second-parameter cluster" may not be strictly applicable, although this 
remains to be established.

The number of known RR Lyrae stars in the NGC~6441 field has been increased 
to 46.  As predicted by Sweigart \& Catelan (1998), the 
periods of the RR Lyrae are unusually long for a cluster of its metallicity,  
a result confirmed and extended in the period-amplitude diagram comparing 
NGC~6441 to other globular clusters.  A few long period RRc stars were also 
found to exist in the cluster, resulting in a smaller than expected gap 
between the longest period RRc and the shortest period RRab stars.  Such 
long-period RRc's are extremely rare among globular clusters, 
with the notable exception of $\omega$~Centauri.

NGC~6441 was found to contain a number of fundamental mode RR Lyrae that are 
both brighter and redder than the other probable RRab found along the 
horizontal branch.  This effect is likely due to blending with unresolved 
red companion stars.  The reddening determined for NGC~6441 is 
$E(B-V)=0.51\pm 0.02$~mag 
with significant differential reddening across the cluster.  From the 
RR Lyrae in the cluster, excluding the brighter and redder RRab stars, the 
mean $V$ magnitude of the horizontal branch was determined to be 
$17.51 \pm 0.02$~mag leading to a range in distance of 10.4 to 11.9~kpc.  

NGC~6441 was shown to stand apart from other Galactic 
globular clusters in that it does not fit in the Oosterhoff classification 
scheme.  The mean periods of the RRab in the cluster are as long as or 
even longer than 
the typical, more metal-poor, Oosterhoff Type II clusters.  This contradiction 
in the trend of increasing period with decreasing metallicity, for a given 
amplitude, implies that the metallicity-luminosity relationship for RR Lyrae 
stars is not universal -- if, in fact, the RR Lyrae stars in NGC~6441 do share 
the cluster's relatively high overall metallicity.

\acknowledgments

This work has been supported by the National Science Foundation under 
grants AST 9528080 and AST 9986943.
Support for M. C. was provided by NASA through Hubble Fellowship 
grant HF-01105.01-98A
awarded by the Space Telescope Science Institute, which is operated 
by the Association of Universities
for Research in Astronomy, Inc., for NASA under contract NAS5-26555. 
A. V. S. gratefully acknowledges support from NASA Astrophysics
Theory Program proposal NRA-99-01-ATP-039.

B. P. would like to thank Peter Stetson for the use of his reduction 
programs and his assistance in getting them to run properly.  Thank 
you to Nancy Silbermann for sharing her knowledge of the reduction programs.  
Thank you to Suzanne Hawley and Tim Beers for their insightful comments.  
Thank you to Brian Sharpee for the use of his spline programs.  We would 
also like to thank Christine Clement for the use of her Fourier 
deconvolution program, and for providing analytical formulae for her 
Oosterhoff lines in the period-amplitude diagram.  We also thank Johanna 
Jurcsik for her assistance in the preliminary Fourier parameters for the 
NGC~6441 variables.  Thank you to the anonymous referee for the insightful 
comments and suggestions.

\clearpage

\figcaption[Pritzl.fig1.ps]{Comparison to the Hesser \& Hartwick (1976) data.  
The differences in magnitude are magnitudes in the present study minus the 
photographic magnitudes in Hesser \& Hartwick (1976). 
\label{fig1}} 

\figcaption[Pritzl.fig2.ps]{Color-magnitude diagrams for the stars located 
in the complete field of view (a), out to a radius of 1.7 arcmin (b) and 
2.7 arcmin (c) from the cluster center, and 6-11 arcmin east of the cluster 
center (d).  \label{fig2}}  

\figcaption[Pritzl.fig3.ps]{Finding charts for the variable stars.  North is 
down and east is left.  \label{fig3}}

\figcaption[Pritzl.fig4.ps]{Color-magnitude diagram for the fundamental 
mode RR Lyrae (filled circles), first overtone RR Lyrae (filled squares), 
and suspected RR Lyrae (filled triangles) in the field of NGC~6441.  
The field RR Lyrae, V45 and V54 in Table~3, are represented by five-pointed 
stars.  
\label{fig4}} 

\figcaption[Pritzl.fig5.ps]{Light curves of variable stars.  \label{fig5}} 

\figcaption[Pritzl.fig6.ps]{Fourier parameter plot using $A_{21}$ vs. 
$\phi_{21}$ to show the distinction between RR Lyrae types.  The filled 
circle represents the field star V54.  \label{fig6}} 

\figcaption[Pritzl.fig7.ps]{Period-amplitude diagram for NGC~6441 in $V$ and 
$B$ showing fundamental mode RR Lyrae (filled squares) and first overtone 
RR Lyrae (open squares).  Fundamental mode RR Lyrae which may be blended 
with companions are denoted by filled circles.  The field star, V54, is 
represented by a five-point star.  An asterisk denotes the probable field 
RR Lyrae, V45.  The line represents the Oosterhoff Type II lines as given 
by Clement (2000; private communication).  \label{fig7}} 

\figcaption[Pritzl.fig8.eps]{Mean period vs. [Fe/H] diagram showing the 
offset of NGC~6441 (square) from the Oosterhoff~I (crosses) and Oosterhoff~II 
(asterisks) globular clusters.  Also plotted is the metal-rich globular cluster 
NGC~6388 (circle) using data from Pritzl et al. (2001).  Data for the 
Oosterhoff type 
I and II clusters are taken from Sandage (1993).  \label{fig8}}  

\figcaption[Pritzl.fig9.ps]{Period distribution histogram for the RR Lyraes 
in NGC~6441.  The dark area is occupied by c-type RR Lyrae variables.  The 
light area is occupied by ab-type RR Lyrae variables.  Only RR Lyrae with 
certain classification are included.  \label{fig9}}

\figcaption[Pritzl.fig10.ps]{Period-amplitude diagram for the 
ab-type RR Lyrae variables of NGC~6441 (filled circles) as compared to field 
RR Lyrae of ${\rm [Fe/H]} \ge -0.8$ (asterisks), V9 in 47 Tuc (open star), 
M3 (open boxes), M15 (filled stars), 
and M68 (triangles).  The smaller filled circles denote variables that are 
believed to be blended with companions.  The open circles represent the 
RRab stars in NGC~6388 from Pritzl et al. (2001).  The boxed area, 
taken from Figure 9 of Layden et al. (1999), denotes the region as 
predicted by Sweigart \& Catelan (1998) where the RR Lyrae 
should be located according to a helium-mixing scenario.
\label{fig10}}

\figcaption[Pritzl.fig11.ps]{Fourier parameter plot using $A_{21}$ vs. 
$\phi_{21}$ to show the distinction between RR Lyrae types in $\omega$ Centauri.  
\label{fig11}} 

\figcaption[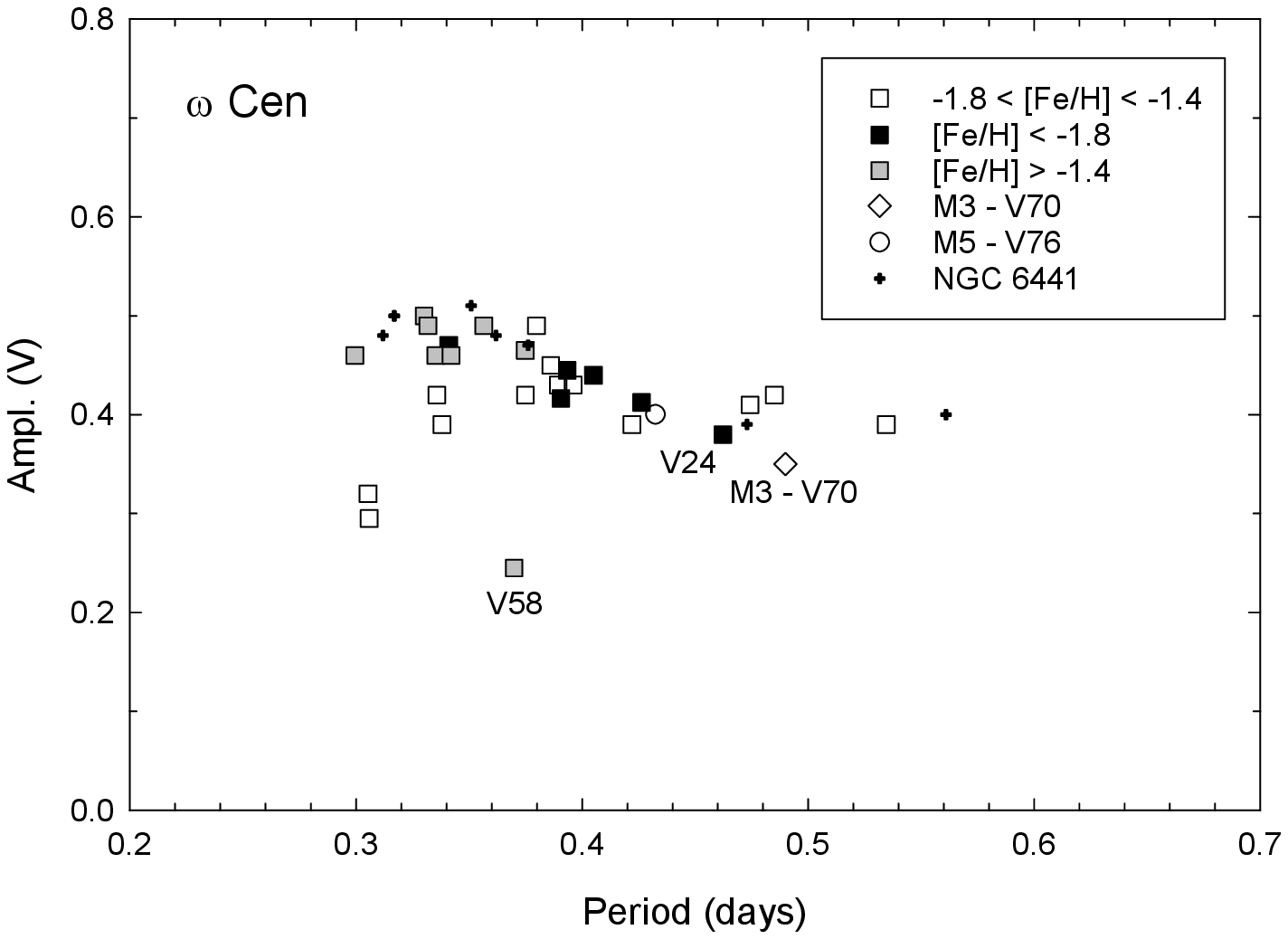]{Period-amplitude diagram for the c-type RR Lyrae 
in $\omega$ Centauri.  \label{fig12}}

\clearpage

\begin{deluxetable}{ccc} 
\tablewidth{0pc}
\footnotesize
\tablecaption{Mean Differences in Photometry\label{tbl-1}}
\tablehead{
\colhead{Reference} & \colhead{$\Delta\,V$} & \colhead{$\Delta\,B$}
          }
\startdata
HST & $0.036\pm0.023$ & $0.003\pm0.033$ \\
Hesser \& Hartwick (photoelectric) & $-0.02\pm0.04$ & $0.02\pm0.04$ \\ 
Hesser \& Hartwick (photographic) & $0.00\pm0.04$ & $0.03\pm0.04$ \\ 
Layden et al. & $-0.385\pm0.011$ & \nodata \\
\enddata
\tablecomments{difference = reference magnitude - magnitude in present study} 
\end{deluxetable}

\clearpage

\begin{deluxetable}{ccccc} 
\tablewidth{0pc}
\footnotesize
\tablecaption{Locations of Variable Stars\label{tbl-2}}
\tablehead{
\colhead{ID} & \colhead{X} & \colhead{Y}
& \colhead{$\Delta\alpha$} & \colhead{$\Delta\delta$} 
          }
\startdata
V1 & 1514.0 & 1161.2 &  49.2 &  -44.3 \\ 
V2 & 1543.2 &  988.5 &  37.5 &   24.2 \\ 
V5 & 1136.0 &  473.8 & 200.1 &  228.5 \\ 
V6 & 1559.6 &  928.4 &  31.0 &   48.1 \\ 
V9 & 1695.5  & 1170.2 & -23.2 &  -47.8 \\ 
V10 & 1441.5 & 1197.1 &  78.1 &  -58.5 \\ 
V37 & 1553.3 &  756.7 &  33.5 &  116.2 \\ 
V38 & 1622.1 &  658.2 &   6.0 &  155.3 \\ 
V39 & 1343.3 &  895.1 & 117.3 &   61.3 \\ 
V40 & 1710.2 & 1189.6 & -29.1 &  -55.5 \\ 
V41 & 1546.0 & 1180.0 &  36.4 &  -51.7 \\ 
V42 & 1637.2 &  817.6 &   0.0 &   92.0 \\ 
V43 & 1523.0 &  911.8 &  45.6 &   54.7 \\ 
V44 & 1598.6 & 1187.9 &  15.4 &  -54.8 \\ 
V45 & 1798.8 & 1390.0 & -64.4 & -135.1 \\ 
V46 & 1996.8 & 1252.3 &-143.5 &  -80.4 \\ 
V47 & 1452.7 & 1809.9 &  73.6 & -301.7 \\ 
V48 &  784.7 & 1212.0 & 340.3 &  -64.4 \\ 
V49 &  994.8 &  813.5 & 256.4 &   93.7 \\ 
V50 & 1458.4 & 1457.3 &  71.4 & -161.8 \\ 
V51 & 1273.2 &  611.8 & 145.3 &  173.7 \\ 
V52 & 1723.7 &  833.7 & -34.4 &   85.6 \\ 
V53 & 1655.6 &  856.5 &  -7.3 &   76.6 \\ 
V54 &  278.2 &  924.9 & 542.5 &   49.5 \\ 
V55 & 1654.9 &  958.7 &  -7.0 &   36.0 \\ 
V56 & 1694.4 &  988.7 & -22.8 &   24.1 \\ 
V57 & 1714.8 & 1028.7 & -30.9 &    8.3 \\ 
V58 & 1586.2 & 1055.8 &  20.3 &   -2.4 \\ 
V59 & 1757.3 & 1119.4 & -47.9 &  -27.6 \\ 
V60 & 1626.7 & 1166.9 &   4.2 &  -46.5 \\ 
V61 & 1585.4 & 1193.7 &  20.7 &  -57.1 \\ 
V62 & 1630.6 & 1203.5 &   2.6 &  -61.0 \\ 
V63 & 1688.0 & 1007.4 & -20.2 &   16.7 \\ 
V64 & 1736.5 & 1057.6 & -39.6 &   -3.1 \\ 
V65 & 1633.2 &  963.3 &   1.6 &   34.2 \\ 
V66 & 1684.5 &  906.7 & -18.8 &   56.7 \\ 
V67 &  710.3 & 1157.3 & 370.0 &  -42.7 \\ 
V68 & 1981.4 & 1553.1 &-137.3 & -199.8 \\ 
V69 & 1162.0 & 1102.9 & 189.7 &  -21.1 \\ 
V70 & 1889.2 & 1181.9 &-100.5 &  -52.5 \\ 
V71 & 1538.4 &  846.8 &  39.4 &   80.4 \\ 
V72 & 1586.1 &  518.2 &  20.4 &  210.9 \\ 
V73 &  283.7 &  635.7 & 540.3 &  164.2 \\ 
V74 & 1469.8 & 1027.1 &  66.8 &    8.9 \\ 
V75 & 1652.0 & 1139.2 &  -5.8 &  -35.5 \\ 
V76 & 1349.6 & 1149.9 & 114.8 &  -39.8 \\ 
V77 & 1130.4 & 1261.8 & 202.3 &  -84.2 \\ 
V78 & 1205.4 & 1629.1 & 172.4 & -229.9 \\ 
V79 & 1607.6 & 1139.4 &  11.8 &  -35.6 \\ 
V80 & 1203.9 & 1288.8 & 172.9 &  -94.9 \\ 
V81 & 1509.9 &  382.2 &  50.8 &  264.9 \\ 
V82 & 1801.4 &  619.8 & -65.4 &  170.5 \\ 
V83 &  866.4 &  743.7 & 307.7 &  121.4 \\ 
V84 & 1753.9 & 1006.6 & -46.5 &   17.0 \\ 
V85 & 1840.5 & 1214.2 & -81.1 &  -65.3 \\ 
V86 & 1880.5 & 1348.5 & -97.1 & -118.6 \\ 
V87 &  285.5 & 1438.1 & 539.6 & -154.1 \\ 
V88 &  592.0 & 1683.6 & 417.2 & -251.6 \\ 
V89 & 1553.9 &  413.2 &  33.2 &  252.5 \\ 
V90 &  483.5 &  943.2 & 460.5 &   42.2 \\ 
V91 & 1625.0 & 1756.8 &   4.9 & -280.6 \\ 
V92 & 1091.9 &  942.3 & 217.7 &   42.5 \\ 
V93 & 1836.7 & 1002.1 & -79.5 &   18.8 \\ 
V94 & 1672.9 &  958.8 & -14.2 &   36.0 \\ 
V95 & 1770.4 & 1222.9 & -53.1 &  -68.7 \\ 
V96 & 1978.3 & 1387.7 &-136.1 & -134.1 \\ 
V97 & 1823.0 & 1099.0 & -74.1 &  -19.6 \\ 
V98 & 1346.7 &  557.4 & 116.0 &  195.3 \\ 
V99 &  185.7 & 1175.1 & 579.4 &  -49.7 \\ 
V100 & 1323.9 &  371.2 & 125.1 &  269.2 \\ 
V101 & 1006.1 &  126.4 & 251.9 &  366.4 \\ 
V102 & 1736.2 & 1215.1 & -39.4 &  -65.6 \\ 
V103 &  849.1 &  757.1 & 314.6 &  116.1 \\ 
V104 & 1928.0 & 1901.9 &-116.0 & -338.2 \\ 
\enddata
\end{deluxetable}

\clearpage

\begin{deluxetable}{ccccccc}
\tablewidth{0pc}
\tablecolumns{7} 
\tablecaption{Mean Properties of RR Lyrae\label{tbl-3}}
\tablehead{
\colhead{ID} & \colhead{Period} & \colhead{$\langle V \rangle$}
& \colhead{$(\bv)_{\rm mag}$} & \colhead{$A_V$} 
& \colhead{$A_B$} & \colhead{Comments}  
          }
\startdata
 V37  & 0.614  & 17.546 & 0.839  & 1.17 & 1.55 & ab \\  
 V38  & 0.735  & 17.462 & 0.857  & 0.77 & 1.07 & ab \\  
 V39  & 0.833  & 17.673 & 0.970  & 0.70 & 0.95 & ab \\ 
 V40  & 0.648  & 17.512 & 0.772  & 1.08 & 1.45 & ab \\  
 V41  & 0.749  & 16.726 & 1.197  & 0.41 & 0.75 & ab \\  
 V42  & 0.813  & 17.475 & 0.900  & 0.58 & 0.80 & ab \\  
 V43  & 0.773  & 17.529 & 0.960  & 0.60 & 0.80 & ab \\  
 V44  & 0.609  & 16.673 & 1.167  & 0.60 & 1.05 & ab \\  
 V45  & 0.503  & 17.375 & 0.814  & 0.87 & 1.07 & ab, Field \\  
 V46  & 0.900  & 17.453 & 0.915  & 0.40 & 0.54 & ab \\  
 V49  & 0.335  & 16.814 & 0.721  & \nodata & \nodata & c?\\  
 V51  & 0.713  & 17.711 & 0.939  & 1.00 & 1.35 & ab, SV8 \\  
 V52  & 0.858  & 17.461 & 0.943  & 0.23 & 0.33 & ab \\  
 V53  & 0.853  & 17.442 & 0.900  & 0.36 & 0.50 & ab \\  
 V54  & 0.620  & 16.535 & 0.929  & 0.51 & 0.67 & ab, Field \\  
 V55  & 0.698  & 17.523 & 0.701  & 0.97 & 1.25 & ab \\  
 V56  & 0.905  & 16.497 & 1.114  & \nodata & 0.64 & ab \\  
 V57  & 0.696  & 17.313 & 0.889  & 0.95 & 1.25 & ab \\  
 V58  & 0.685  & 16.868 & 0.840  & \nodata & 0.70 & ab \\  
 V59  & 0.703  & 17.508 & 0.787  & 0.92 & 1.22 & ab \\  
 V60  & 0.857  & 16.823 & 1.114  & \nodata & 0.29 & ab \\  
 V61  & 0.750  & 17.623 & 0.926  & 0.77 & 1.06 & ab \\  
 V62  & 0.680  & 16.887 & 1.118  & 0.51 & 1.02 & ab \\  
 V63  & 0.700  & 17.064 & 0.755  & \nodata & 0.78 & ab \\  
 V64  & 0.718  & 16.986 & 1.321  & \nodata & 0.95 & ab \\  
 V65  & 0.757  & 16.911 & 1.090  & 0.40 & 0.61 & ab \\  
 V66  & 0.860  & 17.057 & 1.238  & \nodata & 0.44 & ab \\  
 V67  & 0.654  & 16.899 & 0.902  & 0.87 & 1.05 & ab, Field? \\  
 V68  & 0.324  & 16.129 & 0.586  & \nodata & 0.49 & c?, SV1 \\  
 V69  & 0.561  & 17.453 & 0.816  & 0.40 & 0.56 & c, SV2 \\  
 V70  & 0.317  & 17.513 & 0.586  & 0.50 & 0.70 & c, SV4 \\ 
 V71  & 0.362  & 17.471 & 0.726  & 0.48 & 0.65 & c, SV5 \\ 
 V72  & 0.312  & 17.348 & 0.642  & 0.48 & 0.65 & c \\ 
 V73  & 0.320  & 16.968 & 0.865  & 0.38 & 0.52 & c?, Field? \\  
 V74  & 0.317  & 17.578 & 0.705  & 0.50 & 0.65 & c \\ 
 V75  & 0.405  & 17.349 & 0.684  & \nodata & 0.46  & c? \\ 
 V76  & 0.473  & 17.912 & 0.890  & 0.39 & 0.41 & c \\ 
 V77  & 0.376  & 17.494 & 0.679  & 0.47 & 0.65 & c \\ 
 V78  & 0.351  & 17.843 & 0.758  & 0.51 & 0.68 & c \\ 
 V79  & 0.417  & 17.228 & 0.890  & \nodata & 0.68 & c \\ 
 V81  & 0.428  & 17.878 & 0.912  & 0.37 & 0.38 & Binary? \\
 V84  & 0.316  & 17.377 & 0.162  & \nodata & 0.38 & c? \\ 
 V93  & 0.339  & 17.332 & 0.737  & 0.54 & \nodata & c? \\ 
 V94  & 0.386  & 17.368 & 0.801  & 0.36 & 0.54 & c? \\ 
 V95  & 0.090  & 17.611 & 0.652  & 0.55 & 0.67 & $\delta$ Scuti or SX Phe \\ 
 V96  & 0.856  & 17.664 & 0.947  & \nodata & \nodata & ab?\\  
 V97  & 0.844  & 17.446 & 0.951  & \nodata & \nodata & ab? \\ 
V102  & 0.308  & 15.834 & 1.429  & \nodata & 0.32 & c? \\ 
\enddata 
\end{deluxetable}

\clearpage

\begin{deluxetable}{cccccc}
\tablewidth{0pt}
\footnotesize
\tablecaption{Photometry of the Variable Stars (V)\label{tbl-4}}
\tablehead{
\colhead{} & \multicolumn{2}{c}{V37} & & \multicolumn{2}{c}{V38} \\ 
\cline{2-3} \cline{5-6} \\ 
\colhead{HJD-2450000} & \colhead{$V$} & \colhead{$\sigma_{V}$} & & 
 \colhead{$V$} & \colhead{$\sigma_{V}$} 
          }
\startdata
  966.180 &  17.800  &  0.023 &&  17.740  &  0.021 \\ 
  959.158 &  17.136  &  0.018 &&  17.267  &  0.017 \\ 
  959.227 &  17.401  &  0.020 &&  17.378  &  0.018 \\ 
  960.162 &  17.958  &  0.027 &&  17.627  &  0.021 \\ 
  961.327 &  17.874  &  0.022 &&  17.198  &  0.017 \\ 
  961.358 &  17.936  &  0.024 &&  17.261  &  0.018 \\ 
  962.079 &  18.047  &  0.024 &&  17.201  &  0.022 \\ 
  962.125 &  17.460  &  0.021 &&  17.292  &  0.018 \\ 
  962.164 &  16.913  &  0.022 &&  17.351  &  0.017 \\ 
  962.247 &  17.223  &  0.027 &&  17.466  &  0.021 \\ 
  962.285 &  17.357  &  0.021 &&  17.538  &  0.017 \\ 
  962.318 &  17.458  &  0.022 &&  17.558  &  0.019 \\ 
  962.355 &  17.547  &  0.021 &&  17.590  &  0.020 \\ 
  965.085 &  17.996  &  0.021 &&  17.337  &  0.019 \\ 
  965.122 &  18.070  &  0.022 &&  17.399  &  0.020 \\ 
  965.168 &  17.889  &  0.021 &&  17.457  &  0.019 \\ 
  965.206 &  17.191  &  0.018 &&  17.527  &  0.017 \\ 
  965.238 &  16.922  &  0.018 &&  17.558  &  0.018 \\ 
  965.272 &  17.053  &  0.018 &&  17.593  &  0.019 \\ 
  965.303 &  17.189  &  0.020 &&  17.608  &  0.018 \\ 
  965.339 &  17.315  &  0.020 &&  17.648  &  0.020 \\ 
  966.071 &  17.589  &  0.023 &&  17.639  &  0.019 \\ 
  966.103 &  17.649  &  0.020 &&  17.675  &  0.018 \\ 
  966.109 &  17.660  &  0.019 &&  17.696  &  0.017 \\ 
  966.142 &  17.734  &  0.020 &&  17.733  &  0.018 \\ 
  966.211 &  17.859  &  0.022 &&  17.809  &  0.021 \\ 
  966.257 &  17.878  &  0.025 &&  17.844  &  0.021 \\ 
  966.291 &  17.955  &  0.027 &&  17.721  &  0.022 \\ 
  966.319 &  18.002  &  0.025 &&  17.420  &  0.021 \\ 
  966.350 &  18.032  &  0.023 &&  17.296  &  0.021 \\ 
  967.073 &  16.912  &  0.018 &&  17.382  &  0.020 \\ 
  967.104 &  17.024  &  0.018 &&  17.139  &  0.021 \\ 
  967.138 &  17.154  &  0.021 &&  17.058  &  0.020 \\ 
  967.178 &  17.318  &  0.018 &&  17.124  &  0.017 \\ 
  967.251 &  17.515  &  0.019 &&  17.260  &  0.018 \\ 
  967.285 &  17.588  &  0.022 &&  17.326  &  0.019 \\ 
  967.318 &  17.622  &  0.021 &&  17.370  &  0.018 \\ 
  967.352 &  17.681  &  0.021 &&  17.425  &  0.020 \\ 
  968.063 &  17.861  &  0.027 &&  17.416  &  0.020 \\ 
  968.097 &  17.879  &  0.020 &&  17.446  &  0.018 \\ 
  968.140 &  17.966  &  0.021 &&  17.511  &  0.019 \\ 
  968.170 &  18.011  &  0.024 &&  17.547  &  0.019 \\ 
  968.220 &  18.009  &  0.021 &&  17.592  &  0.021 \\ 
  968.264 &  17.377  &  0.018 &&  17.630  &  0.020 \\ 
  968.298 &  16.886  &  0.018 &&  17.688  &  0.018 \\ 
\enddata
\tablecomments{The complete version of this table is in the electronic 
edition of the Journal.  The printed edition contains only a sample.}
\end{deluxetable}

\clearpage

\begin{deluxetable}{cccccc}
\tablewidth{0pt}
\footnotesize
\tablecaption{Photometry of the Variable Stars (B)\label{tbl-5}}
\tablehead{
\colhead{} & \multicolumn{2}{c}{V37} & & \multicolumn{2}{c}{V38} \\ 
\cline{2-3} \cline{5-6} \\ 
\colhead{HJD-2450000} & \colhead{$B$} & \colhead{$\sigma_{B}$} & & 
 \colhead{$B$} & \colhead{$\sigma_{B}$} 
          }
\startdata
  966.188 &  18.801  &  0.017 &&  18.720  &  0.019 \\
  959.146 &  17.702  &  0.010 &&  17.977  &  0.015 \\
  960.147 &  18.887  &  0.072 &&  18.583  &  0.039 \\
  961.318 &  18.801  &  0.018 &&  17.891  &  0.009 \\
  961.350 &  18.874  &  0.027 &&  17.954  &  0.018 \\
  962.086 &  19.003  &  0.034 &&  17.986  &  0.025 \\
  962.133 &  \nodata  &  \nodata &&  \nodata  &  \nodata \\
  962.172 &  17.490  &  0.011 &&  18.192  &  0.014 \\
  962.239 &  17.854  &  0.019 &&  18.345  &  0.018 \\
  962.276 &  18.072  &  0.019 &&  18.420  &  0.018 \\
  962.310 &  18.241  &  0.017 &&  18.458  &  0.016 \\
  962.346 &  18.380  &  0.018 &&  18.522  &  0.019 \\
  965.093 &  18.986  &  0.031 &&  18.170  &  0.015 \\
  965.129 &  19.018  &  0.020 &&  18.238  &  0.019 \\
  965.161 &  18.913  &  0.015 &&  18.313  &  0.013 \\
  965.199 &  18.110  &  0.010 &&  18.392  &  0.008 \\
  965.231 &  17.489  &  0.010 &&  18.468  &  0.014 \\
  965.264 &  17.613  &  0.013 &&  18.530  &  0.016 \\
  965.296 &  17.802  &  0.017 &&  18.518  &  0.014 \\
  965.328 &  18.019  &  0.017 &&  18.567  &  0.013 \\
  966.079 &  18.501  &  0.025 &&  18.584  &  0.021 \\
  966.119 &  18.568  &  0.021 &&  18.640  &  0.018 \\
  966.149 &  18.642  &  0.015 &&  18.693  &  0.014 \\
  966.218 &  18.819  &  0.012 &&  18.798  &  0.015 \\
  966.250 &  18.803  &  0.018 &&  18.792  &  0.018 \\
  966.281 &  18.909  &  0.027 &&  18.705  &  0.020 \\
  966.312 &  18.960  &  0.024 &&  18.367  &  0.020 \\
  966.343 &  19.028  &  0.024 &&  18.139  &  0.018 \\
  967.065 &  17.496  &  0.021 &&  18.195  &  0.027 \\
  967.097 &  17.594  &  0.017 &&  17.914  &  0.023 \\
  967.130 &  17.769  &  0.016 &&  17.724  &  0.018 \\
  967.170 &  18.010  &  0.014 &&  17.788  &  0.014 \\
  967.203 &  18.188  &  0.017 &&  17.914  &  0.011 \\
  967.258 &  18.356  &  0.014 &&  18.046  &  0.013 \\
  967.292 &  18.443  &  0.017 &&  18.148  &  0.013 \\
  967.325 &  18.518  &  0.017 &&  18.209  &  0.013 \\
  967.360 &  18.598  &  0.013 &&  18.302  &  0.019 \\
  968.071 &  18.783  &  0.042 &&  18.263  &  0.027 \\
  968.105 &  18.853  &  0.038 &&  18.316  &  0.017 \\
  968.147 &  18.986  &  0.047 &&  18.391  &  0.022 \\
  968.178 &  19.068  &  0.039 &&  18.471  &  0.027 \\
  968.212 &  18.990  &  0.026 &&  18.541  &  0.023 \\
  968.256 &  18.278  &  0.015 &&  18.554  &  0.017 \\
  968.290 &  17.501  &  0.011 &&  18.618  &  0.019 \\
\enddata 
\tablecomments{The complete version of this table is in the electronic 
edition of the Journal.  The printed edition contains only a sample.}
\end{deluxetable}

\clearpage

\begin{deluxetable}{ccccccccc}
\tablewidth{0pt}
\footnotesize
\tablecaption{Fourier Coefficients For RR Lyrae\label{tbl-6}}
\tablehead{
\colhead{ID} & \colhead{$A_{\rm 1}$} & \colhead{$A_{\rm 21}$} & 
\colhead{$A_{\rm 31}$} & \colhead{$A_{\rm 41}$} & \colhead{$\phi_{\rm 21}$} & 
\colhead{$\phi_{\rm 31}$} & \colhead{$\phi_{\rm 41}$} & \colhead{$D_m$}  
          }
\startdata
 V37 & 0.397 & 0.525 & 0.320 & 0.187 & 4.330 & 2.362$\pm$0.047 & 0.634 & 1.52 \\ 
 V38 & 0.293 & 0.472 & 0.237 & 0.114 & 4.418 & 2.708$\pm$0.073 & 0.960 & 5.79 \\ 
 V39 & 0.282 & 0.424 & 0.154 & 0.071 & 4.604 & 2.952$\pm$0.117 & 1.486 & 9.21 \\ 
 V40 & 0.390 & 0.515 & 0.348 & 0.152 & 4.335 & 2.452$\pm$0.046 & 0.655 & 6.56 \\ 
 V42 & 0.239 & 0.414 & 0.118 & 0.053 & 4.658 & 2.903$\pm$0.101 & 1.341 & 6.45 \\ 
 V43 & 0.237 & 0.421 & 0.188 & 0.059 & 4.377 & 2.843$\pm$0.088 & 1.070 & 8.30 \\ 
 V49 & 0.180 & 0.074 & 0.037 & 0.027 & 5.071 & 3.402$\pm$0.236 & 2.483 & \nodata \\ 
 V51 & 0.360 & 0.509 & 0.291 & 0.147 & 4.391 & 2.842$\pm$0.062 & 0.913 & 8.15 \\ 
 V55 & 0.368 & 0.475 & 0.239 & 0.101 & 4.527 & 2.808$\pm$0.129 & 0.857 & 8.33 \\ 
 V57 & 0.326 & 0.529 & 0.330 & 0.156 & 4.370 & 2.623$\pm$0.083 & 0.695 & 2.72 \\ 
 V59 & 0.331 & 0.432 & 0.273 & 0.144 & 4.454 & 2.762$\pm$0.160 & 0.849 & 8.20 \\ 
 V61 & 0.285 & 0.460 & 0.200 & 0.095 & 4.473 & 2.811$\pm$0.093 & 1.359 & 7.24 \\ 
 V62 & 0.168 & 0.600 & 0.337 & 0.150 & 4.193 & 2.344$\pm$0.119 & 0.562 & 4.29 \\ 
 V65 & 0.124 & 0.550 & 0.267 & 0.083 & 4.269 & 2.353$\pm$0.135 & 0.041 & 8.04 \\ 
 V67 & 0.316 & 0.405 & 0.282 & 0.176 & 4.112 & 2.105$\pm$0.279 & 0.186 & 4.06 \\ 
 V69 & 0.178 & 0.175 & 0.157 & 0.052 & 3.037 & 0.353$\pm$0.115 & 5.028 & \nodata \\ 
 V70 & 0.264 & 0.096 & 0.052 & 0.041 & 4.185 & 3.561$\pm$0.488 & 3.525 & \nodata \\ 
 V71 & 0.226 & 0.015 & 0.072 & 0.045 & 4.094 & 5.443$\pm$0.235 & 4.050 & \nodata \\ 
 V72 & 0.244 & 0.105 & 0.036 & 0.024 & 4.097 & 4.055$\pm$0.464 & 1.889 & \nodata \\ 
 V73 & 0.203 & 0.108 & 0.091 & 0.036 & 5.102 & 4.054$\pm$0.202 & 3.155 & \nodata \\ 
 V74 & 0.242 & 0.088 & 0.070 & 0.038 & 3.845 & 4.478$\pm$0.271 & 3.258 & \nodata \\ 
 V76 & 0.247 & 0.235 & 0.316 & 0.206 & 0.464 & 0.053$\pm$0.059 & 0.017 & \nodata \\ 
 V77 & 0.233 & 0.112 & 0.052 & 0.035 & 3.526 & 5.579$\pm$0.325 & 4.260 & \nodata \\ 
 V78 & 0.260 & 0.083 & 0.034 & 0.049 & 4.023 & 4.959$\pm$0.458 & 3.201 & \nodata \\
 V79 & 0.211 & 0.158 & 0.197 & 0.123 & 4.181 & 5.063$\pm$0.450 & 0.161 & \nodata \\
 V93 & 0.223 & 0.157 & 0.051 & 0.048 & 3.822 & 2.349$\pm$0.578 & 4.923 & \nodata \\ 
 V94 & 0.183 & 0.051 & 0.060 & 0.055 & 0.051 & 5.245$\pm$0.671 & 5.203 & \nodata \\ 
\enddata
\end{deluxetable}

\clearpage

\begin{deluxetable}{cccccccc}
\tablewidth{0pt}
\footnotesize
\tablecaption{Fourier Coefficients For Binary Stars and Other RR 
Lyrae\label{tbl-7}}
\tablehead{
\colhead{ID} & \colhead{$A_{\rm 1}$} & \colhead{$A_{\rm 21}$} & 
\colhead{$A_{\rm 31}$} & \colhead{$A_{\rm 41}$} & \colhead{$\phi_{\rm 21}$} & 
\colhead{$\phi_{\rm 31}$} & \colhead{$\phi_{\rm 41}$} 
          }
\startdata
V41 & 0.153 & 0.519 & 0.198 & 0.201 & 4.114 & 2.358$\pm$0.126 & 0.464 \\
V44 & 0.187 & 0.611 & 0.359 & 0.196 & 4.289 & 2.002$\pm$0.112 & 0.470 \\ 
V45 & 5.726 & 0.816 & 0.605 & 0.414 & 6.202 & 6.200$\pm$0.003 & 6.166 \\
V46 & 0.162 & 0.315 & 0.100 & 0.026 & 4.520 & 2.506$\pm$0.385 & 1.019 \\
V47 & 0.197 & 1.132 & 0.792 & 0.753 & 6.093 & 6.033$\pm$0.139 & 5.894 \\
V48 & 0.020 & 6.562 & 0.387 & 1.034 & 5.914 & 5.557$\pm$0.309 & 5.645 \\
V50 & 0.007 & 32.80 & 1.772 & 10.01 & 5.301 & 2.999$\pm$3.482 & 4.283 \\
V52 & 0.090 & 0.615 & 0.328 & 0.259 & 5.236 & 6.084$\pm$0.116 & 0.818 \\
V53 & 0.185 & 0.273 & 0.141 & 0.120 & 5.560 & 0.204$\pm$0.141 & 0.530 \\
V54 & 0.225 & 0.328 & 0.135 & 0.039 & 4.078 & 1.992$\pm$0.059 & 6.113 \\
V56 & 0.126 & 0.223 & 0.303 & 0.237 & 3.141 & 4.116$\pm$0.418 & 6.055 \\
V58 & 0.254 & 0.476 & 0.107 & 0.015 & 3.849 & 2.229$\pm$0.993 & 2.525 \\ 
V60 & 0.128 & 0.127 & 0.300 & 0.232 & 6.038 & 1.554$\pm$0.170 & 1.681 \\
V63 & 0.174 & 0.516 & 0.343 & 0.055 & 4.867 & 3.170$\pm$0.357 & 0.222 \\
V64 & 0.114 & 0.557 & 0.213 & 0.139 & 4.279 & 2.290$\pm$0.171 & 0.508 \\
V66 & 0.085 & 0.917 & 0.422 & 0.387 & 5.084 & 5.525$\pm$0.273 & 0.205 \\
V68 & 0.231 & 0.127 & 0.142 & 0.046 & 4.009 & 4.451$\pm$0.448 & 4.510 \\
V75 & 0.155 & 0.125 & 0.029 & 0.056 & 3.423 & 5.688$\pm$1.859 & 1.072 \\
V80 & 0.010 & 17.32 & 1.704 & 2.400 & 5.549 & 4.895$\pm$0.599 & 4.644 \\
V81 & 0.155 & 0.278 & 0.087 & 0.048 & 0.054 & 0.018$\pm$0.277 & 2.684 \\
V82 & 0.032 & 2.850 & 0.361 & 0.327 & 0.064 & 0.270$\pm$0.337 & 0.002 \\
V83 & 0.021 & 5.221 & 0.580 & 0.657 & 0.115 & 0.438$\pm$0.455 & 0.205 \\
V84 & 0.204 & 0.072 & 0.063 & 0.028 & 4.127 & 5.175$\pm$0.862 & 5.770 \\
V85 & 0.028 & 5.349 & 0.414 & 0.659 & 0.537 & 0.982$\pm$0.605 & 0.943 \\
V86 & 0.039 & 5.102 & 0.335 & 1.368 & 5.637 & 4.642$\pm$0.738 & 5.070 \\
V87 & 0.031 & 3.102 & 0.141 & 0.343 & 2.420 & 3.393$\pm$0.539 & 5.017 \\
V88 & 0.006 & 33.81 & 1.239 & 6.771 & 0.143 & 0.698$\pm$2.327 & 0.335 \\
V89 & 0.030 & 4.511 & 0.173 & 0.691 & 5.505 & 5.406$\pm$0.911 & 5.484 \\
V90 & 0.014 & 9.963 & 1.222 & 1.090 & 0.781 & 0.796$\pm$1.017 & 1.007 \\
V91 & 0.021 & 13.64 & 1.588 & 2.989 & 0.840 & 0.938$\pm$2.154 & 1.625 \\
V92 & 0.037 & 3.780 & 0.337 & 0.903 & 5.807 & 0.469$\pm$0.673 & 5.349 \\
V95 & 0.209 & 0.283 & 0.075 & 0.078 & 3.951 & 5.885$\pm$0.594 & 4.782 \\
V96 & 0.063 & 1.018 & 1.092 & 0.685 & 4.981 & 3.030$\pm$0.442 & 0.751 \\
V97 & 0.126 & 1.054 & 1.056 & 0.884 & 3.960 & 1.276$\pm$0.160 & 4.783 \\
V100 & 0.285 & 0.953 & 0.429 & 0.510 & 5.837 & 5.673$\pm$0.220 & 6.148 \\
V102 & 0.042 & 0.287 & 0.056 & 0.144 & 4.737 & 5.587$\pm$1.310 & 2.277 \\
V103 & 0.037 & 2.534 & 0.152 & 0.315 & 5.705 & 5.472$\pm$0.825 & 4.870 \\
V104 & 0.045 & 6.661 & 0.117 & 3.180 & 2.064 & 0.224$\pm$5.332 & 4.437 \\
\enddata
\end{deluxetable}

\clearpage

\begin{deluxetable}{cccc}
\tablewidth{0pt}
\footnotesize
\tablecaption{Fourier Coefficients For Alternate Periods\label{tbl-8}}
\tablehead{
\colhead{ID} & \colhead{Period} & \colhead{$A_{\rm 21}$} & 
\colhead{$\phi_{\rm 21}$} 
          }
\startdata 
V86 & 0.162 & 0.231 & 0.171 \\ 
V88 & 0.226 & 0.225 & 0.053 \\ 
V49 & 0.335 & 0.074 & 5.071 \\ 
V81 & 0.428 & 0.228 & 0.054 \\ 
\enddata 
\end{deluxetable}

\clearpage

\begin{deluxetable}{cccc}
\tablewidth{0pc}
\footnotesize
\tablecaption{Reddening Determinations\label{tbl-9}}
\tablehead{
\colhead{} & \multicolumn{2}{c}{$E(\bv)$} \\ 
\cline{2-3} \\ 
\colhead{ID} & \colhead{Pritzl et al.} & \colhead{Layden et al.} & 
\colhead{Comments} 
          }
\startdata
V37 & 0.528 & 0.410  \\ 
V38 & 0.499 & 0.412  \\ 
V39 & 0.570 & 0.548  \\ 
V40 & 0.440 & 0.468  \\ 
V41 & 0.822 & 0.683 & Bright \& Red \\ 
V42 & 0.484 & 0.444  \\ 
V43 & 0.504 & 0.413  \\ 
V44 & 0.847 & 0.676 & Bright \& Red \\ 
V51 & 0.600 & \nodata \\ 
V54 & 0.541 & \nodata & Field  \\ 
V57 & 0.561 & \nodata \\ 
V59 & 0.427 & \nodata \\ 
V61 & 0.535 & \nodata \\ 
V62 & 0.822 & \nodata & Bright \& Red \\  

\enddata
\end{deluxetable}

\clearpage

\begin{deluxetable}{ccccccc}
\tablewidth{0pc}
\footnotesize
\tablecaption{Mean Properties of Binary Stars\label{tbl-10}}
\tablehead{
\colhead{ID} & \colhead{Period} & \colhead{$\langle V \rangle$}
& \colhead{$(\bv)_{\rm mag}$} & \colhead{$A_V$} & 
\colhead{$A_B$} & \colhead{Comments}
          }
\startdata
 V47 & 0.703 & 16.489 & 0.972 & 1.20 & 1.10 & Detached \\ 
 V48 & 0.668 & 15.501 & 0.864 & 0.31 & 0.32 & Contact \\ 
 V50 & 0.433 & 18.183 & 1.138 & 0.52 & 0.60 & Contact \\ 
 V80 & 0.900 & 17.409 & 0.966 & 0.38 & 0.40 & Contact, SV3 \\ 
 V82 & 0.747 & 16.505 & 0.912 & 0.24 & 0.25 & Contact \\ 
 V83 & 0.622 & 17.625 & 0.948 & 0.28 & 0.25 & Contact \\ 
 V85 & 0.283 & 17.636 & 1.239 & 0.36 & 0.41 & Contact \\ 
 V86 & 0.325 & 17.949 & 1.090 & 0.48 & 0.50 & Contact \\ 
 V87 & 0.369 & 17.140 & 1.117 & 0.24 & 0.26 & Contact \\ 
 V88 & 0.452 & 17.698 & 1.122 & 0.44 & 0.46 & Contact \\ 
 V89 & 0.456 & 18.486 & 0.836 & 0.34 & 0.34 & Contact \\ 
 V90 & 0.726 & 18.523 & 0.973 & 0.30 & 0.34 & Contact \\ 
 V91 & 0.457 & 19.518 & 1.103 & 0.70 & 0.70 & Contact \\ 
 V92 & 0.577 & 18.581 & 0.945 & 0.34 & 0.39 & Contact \\ 
 V100 & 1.66 & 16.966 & 0.846 & 1.05 & 1.15 & Detached \\ 
 V101 & 3.50 & 18.319 & 1.204 & 1.75 & 1.80 & Detached \\ 
 V103 & 0.673 & 18.411 & 0.931 & 0.23 & 0.25 & Contact \\ 
 V104 & 0.735 & 19.407 & 1.243 & 1.05 & 1.15 & Contact \\  
\enddata
\end{deluxetable}

\clearpage 
\begin{deluxetable}{cccccccc} 
\tablewidth{0pc}
\footnotesize
\tablecaption{Mean Properties of Suspected Variables\label{tbl-11}}  
\tablehead{
\colhead{ID} & \colhead{Period} & \colhead{$\langle V \rangle$} & 
\colhead{$(\bv)_{\rm mag}$} & \colhead{$A_V$} & 
\colhead{$A_B$} & \colhead{$\Delta\alpha$} & \colhead{$\Delta\delta$}  
          }
\startdata
SV10 & 0.376 & 17.666 & 1.028 & 0.10 & 0.11 & -29.4 & 169.8 \\ 
SV11 & 0.474 & 17.707 & 1.051 & 0.14 & 0.15 & 388.5 & 206.0 \\ 
SV12 & 0.860 & 17.572 & 0.862 & 0.10 & 0.12 & -11.1 & -72.8 \\ 
SV13 & 1.11  & 16.548 & 1.483 & 0.17 & 0.17 & 136.9 & 323.5 \\ 
SV14 & 0.55  & 17.348 & 0.891 & 0.30 & 0.46 & -28.7 & -21.3 \\ 
SV15 & 0.61  & 16.808 & 0.892 & 0.40 & 0.40 & -26.4 & -0.9  \\ 
\enddata
\end{deluxetable}

\clearpage

\begin{deluxetable}{cccccc}
\tablewidth{0pc}
\footnotesize
\tablecaption{Cluster properties\label{tbl-12}}
\tablehead{
\colhead{Cluster} & \colhead{Type} & \colhead{[Fe/H]}
& \colhead{$\langle P_{\rm ab} \rangle$} & \colhead{$\langle P_{\rm c} 
\rangle$} & \colhead{$N_{\rm c}/N_{\rm RR}$}
          }
\startdata
M3        & Oo~I  & $-1.6$ & 0.56 & 0.32 & 0.16\\
M15       & Oo~II & $-2.2$ & 0.64 & 0.38 & 0.48\\
NGC~6441  &   ?   & $-0.5$ & 0.75 & 0.38 & 0.31\\
\enddata
\end{deluxetable}

\clearpage

\begin{deluxetable}{ccccc}
\tablewidth{0pc}
\footnotesize
\tablecaption{RRc Parameters\label{tbl-13}}
\tablehead{
\colhead{ID} & \colhead{$M/M_{\odot}$} & 
\colhead{$\log\,(L/L_{\odot})$} & \colhead{$T_{\rm eff}$} & 
\colhead{$M_V$}
          }
\startdata 
V70  & 0.53 & 1.68 & 7388 & 0.79 \\ 
V71  & 0.36 & 1.64 & 7435 & 0.76 \\ 
V72  & 0.48 & 1.65 & 7431 & 0.82 \\ 
V73  & 0.49 & 1.66 & 7388 & 0.77 \\ 
V74  & 0.43 & 1.63 & 7458 & 0.82 \\ 
V77  & 0.36 & 1.64 & 7351 & 0.78 \\ 
V78  & 0.41 & 1.65 & 7374 & 0.76 \\ 
Mean & $0.47\pm0.05$ & $1.65\pm0.02$ & $7408\pm35$ & $0.79\pm0.03$ \\ 
\enddata 
\tablecomments{The mean was calculated excluding V71 and V77.} 
\end{deluxetable}

\clearpage 

\begin{deluxetable}{cccccc}
\tablewidth{0pc}
\footnotesize
\tablecaption{RRab Parameters\label{tbl-14}}
\tablehead{ 
\colhead{ID} & \colhead{$M/M_{\odot}$} & 
\colhead{$\log\,(L/L_{\odot})$} & \colhead{$\log\,T_{\rm eff}$} & 
\colhead{$M_{\rm V}$} & \colhead{[Fe/H]}  
          }
\startdata
V37 & 0.54 & 1.63 & 3.83 & 0.74 & -0.95 \\
V38 & 0.55 & 1.69 & 3.81 & 0.66 & -1.14 \\
V40 & 0.54 & 1.66 & 3.82 & 0.71 & -1.01 \\
V43 & 0.55 & 1.69 & 3.81 & 0.65 & -1.16 \\
V51 & 0.53 & 1.65 & 3.82 & 0.67 & -0.84 \\
V55 & 0.53 & 1.65 & 3.82 & 0.68 & -0.80 \\
V57 & 0.55 & 1.66 & 3.82 & 0.69 & -1.04 \\
V59 & 0.53 & 1.65 & 3.82 & 0.69 & -0.89 \\
V61 & 0.55 & 1.66 & 3.81 & 0.65 & -1.08 \\
Mean & 0.54 & 1.66 & 3.82 & 0.68 & -0.99 \\
\enddata
\end{deluxetable}

\clearpage 

\begin{deluxetable}{ccl}
\tablewidth{0pc}
\footnotesize
\tablecaption{M15 Compatibility Condition\label{tbl-14}}
\tablehead{ 
\colhead{ID} & \colhead{$D_m$} & \colhead{Light Curve Source}
          }
\startdata
02 & 3.580 & Silbermann \& Smith (1995) \\ 
09 & 4.795 & Silbermann \& Smith (1995) \\ 
08 & 4.423 & Bingham et al. (1984) \\ 
12 & 5.500 & Bingham et al. (1984) \\ 
25 & 5.628 & Bingham et al. (1984) \\ 
\enddata
\end{deluxetable}

\end{document}